\documentclass[
nofootinbib, 
amsmath,
amssymb,
superscriptaddress,
prd
]{revtex4-2}

\usepackage{slashed}
\usepackage{graphicx}
\usepackage{dcolumn}
\usepackage{bm}
\usepackage{hyperref}
\usepackage{tcolorbox}
\usepackage[mathlines]{lineno}

\usepackage{ulem}
\usepackage{booktabs, multirow,tabularx,makecell}
\usepackage{enumitem}
\usepackage{todonotes}
\usepackage{slashed}
\usepackage{amsmath}
\usepackage{booktabs,tabularx,ragged2e,enumitem,mathtools,bm}
\usepackage[percent]{overpic}
\newcolumntype{Y}{>{\RaggedRight\arraybackslash}X}

\setlist[enumerate]{leftmargin=1.2em,itemsep=2pt,topsep=2pt,parsep=0pt,partopsep=0pt}
\allowdisplaybreaks
\newcommand{\na}{\notag\\} 

\newcommand{\dhd}{{\textstyle d} \lower.03ex\hbox{\kern-0.38em$^{\scriptstyle-}$}\kern-0.05em{}}
\usepackage{xcolor}
\definecolor{revieworange}{RGB}{230,126,34}

\usepackage{subcaption}
\usepackage[framemethod=tikz]{mdframed}

\mdfdefinestyle{proofstyle}{%
    backgroundcolor=gray!20, 
    linecolor=gray,          
    outerlinewidth=1pt,
    roundcorner=5pt,
    skipabove=10pt,
    skipbelow=10pt
}

\newmdtheoremenv[style=proofstyle]{proofenv}{Details}

\newcommand{\dd}{\mathrm d}

\newcommand{\cP}{\mathcal P}

\DeclareRobustCommand{\xperp}{\bm{x}_{\perp}}
\DeclareRobustCommand{\yperp}{\bm{y}_{\perp}}
\DeclareRobustCommand{\zperp}{\bm{z}_{\perp}}
\DeclareRobustCommand{\zoneperp}{\bm{z}_{1\perp}}
\DeclareRobustCommand{\ztwoperp}{\bm{z}_{2\perp}}
\DeclareRobustCommand{\rperp}{\bm{r}_{\perp}}
\DeclareRobustCommand{\bperp}{\bm{b}_{\perp}}
\DeclareRobustCommand{\wperp}{\bm{w}_{\perp}}
\DeclareRobustCommand{\zeroperp}{\bm{0}_{\perp}}
\DeclareRobustCommand{\hatxperp}{\bm{\hat{x}}_{\perp}}
\DeclareRobustCommand{\zetaperp}[1]{\bm{\zeta}_{#1}}
\DeclareRobustCommand{\hatzetaperp}[1]{\bm{\hat{\zeta}}_{#1}}
\DeclareRobustCommand{\kperp}{\bm{k}_{\perp}}
\DeclareRobustCommand{\koneperp}{\bm{k}_{1\perp}}
\DeclareRobustCommand{\ktwoperp}{\bm{k}_{2\perp}}
\DeclareRobustCommand{\pperp}{\bm{p}_{\perp}}
\DeclareRobustCommand{\Pperp}{\bm{P}_{\perp}}
\DeclareRobustCommand{\qperp}{\bm{q}_{\perp}}
\DeclareRobustCommand{\kqperp}{\bm{k}_{q\perp}}
\DeclareRobustCommand{\bDeltaPerp}{\bm{\Delta}_{\perp}}
\DeclareRobustCommand{\vperp}{\bm{v}_{\perp}}

\newcommand{\WL}[3]{[#1,#2]_{#3}}
\DeclareRobustCommand{\Ptop}{{\mathrm P}_\perp}
\DeclareRobustCommand{\Pop}{{\mathrm P}}
\begin{document}
\title{DIS Dijet Production: An operator basis bridging eikonal and TMD regimes}
\author{Tiyasa~Kar}
\affiliation{Department of Physics and Astronomy, North Carolina State University, Raleigh, NC 27695, USA}
\author{Swagato~Mukherjee}
\affiliation{Physics Department, Brookhaven National Laboratory, Upton, New York 11973, USA}
\author{Vladimir~V.~Skokov}
\affiliation{Department of Physics and Astronomy, North Carolina State University, Raleigh, NC 27695, USA}
\author{Shaswat~Tiwari}
\email{stiwari1@bnl.gov}
\affiliation{Physics Department, Brookhaven National Laboratory, Upton, New York 11973, USA}
\author{Fei~Yao}
\email{fyao@bnl.gov}
\affiliation{Physics Department, Brookhaven National Laboratory, Upton, New York 11973, USA}
\begin{abstract}
We propose an operator basis for the {unpolarized} deep-inelastic scattering (DIS) quark--antiquark dijet production process smoothly  connecting the all-twist eikonal regime to the back-to-back leading-twist TMD regime at arbitrary Bjorken $x$. Starting from the background-field quark propagator we construct an operator basis that organizes the eikonal and twist expansions within a unified formulation. Utilizing this operator basis we derive the dijet production amplitudes retaining all contributions required by either the leading-eikonal or the leading-twist description. In the eikonal limit the resulting amplitudes reproduce the all-twist Color Glass Condensate result, while in the back-to-back limit they reduce to the leading-twist gluon TMD-based result at arbitrary $x$. The operator basis can be systematically extended to include sub-eikonal and higher-twist corrections. Further, by examining the relationship of this operator basis to the improved TMD (iTMD) factorization we recover the iTMD structure in the eikonal limit. However, we find at non-zero $x$ the transverse resummation generates longitudinal phases that prevent factorization in terms of a conventional TMD operator.

\end{abstract}
%
\date{\today}
\maketitle
%
\section{Introduction}
\label{sec:intro}
Correlations between two jets in deep-inelastic scattering (DIS) provide direct access to the transverse-momentum structure of gluons in hadrons and nuclei. In the back-to-back regime, where the dijet imbalance is much smaller than the individual jet momenta, the cross section admits a twist expansion in terms of transverse-momentum-dependent (TMD) operators \cite{delCastillo:2020omr}. In the eikonal ($x\to0$) limit, the same process can be treated in the Color Glass Condensate (CGC) effective theory \cite{McLerran:1993ni,McLerran:1993ka,Gelis:2010nm}, which resums multiple scattering from a dense gluon background. This admits a description in terms of all-twist operators in an expansion in eikonality \cite{Dominguez:2011wm,Dominguez:2011br,Metz:2011wb,Dumitru:2015gaa,Caucal:2023nci,Caucal:2023fsf}. The connections between these operator descriptions make dijet correlations a particularly useful observable for the future Electron--Ion Collider (EIC) program \cite{Dumitru:2018kuw,Mantysaari:2019hkq,AbdulKhalek:2021gbh}.

These two limiting descriptions of DIS dijet production are well established in terms of operators in the Hilbert space of the target background field. Perturbative corrections beyond leading order have been worked out for
back-to-back dijet production at small $x$
\cite{CaucalEtAl2021DijetNLO,CaucalEtAl2022BackToBackNLO,
Caucal:2023nci,Caucal:2023fsf}. In the high-energy limit, the eikonal CGC description resums multiple scattering through Wilson-line operators \cite{Dominguez:2011wm,Dominguez:2011br,Caucal:2023nci,Caucal:2023fsf}. In the back-to-back regime the process admits a systematic twist expansion in terms of gluon TMD operators, with results available at leading twist and, more recently, at twist-three accuracy
\cite{delCastillo:2020omr,Mukherjee:2026cte,Mukherjee:2026six}. Although these descriptions apply to different kinematic limits, they describe the same underlying scattering process. It is therefore natural to ask whether they can be embedded into a unifying operator basis that systematically captures both limits.

Several approaches have been developed to go beyond these limiting descriptions. Sub-eikonal corrections associated with the finite longitudinal extent of the target and subleading background-field components have been studied in Refs.~\cite{Altinoluk:2022jkk,Altinoluk:2024zom,Agostini:2024xqs,
Armesto:2026qwh,Altinoluk:2026gdc}. The relation between the high-energy and twist descriptions has also been investigated in Refs.~\cite{BoussarieMehtarTani2022NovelUGD, BoussarieMehtarTani2022PartialTwist, BoussarieMehtarTani2024ComptonPTE,Kar:2026vzk}. These developments motivate a more systematic question: whether the eikonal and twist expansions can be organized within a common operator basis that reproduces the established results in both limits and can be extended systematically beyond them.

To address this question we investigate whether such a unifying operator basis can be constructed starting from the 
quark propagator in a background gluon field. Our investigation leads to a new organization of the background-field propagator in terms of a semi-infinite dipole operator, with corrections entering as field-strength insertions into the Wilson lines. We then derive the {unpolarized DIS} dijet production amplitude in the gluon background for both longitudinally and transversely polarized photons in terms of these operators for arbitrary values of $x$. This allows us to study, reproduce and connect the eikonal and TMD limits of DIS dijet production.

Moreover, we show how our framework naturally connects to the improved-TMD (iTMD) framework. The iTMD framework was introduced to extend the small-$x$ TMD description away from the strict back-to-back limit \cite{Kotko:2015ura}. It resums kinematic powers of the dijet imbalance over the hard momentum while retaining the gauge-link structure of the small-$x$ gluon distributions. The framework has been developed and applied to dijet, trijet, heavy-quark-pair, and photon--jet production in dilute--dense collisions
\cite{Altinoluk:2019wyu,Bury:2020ndc,Altinoluk:2021ygv,Fujii:2020bkl,Kutak:2021kaw,Ganguli:2023joy}. Comparisons with the full CGC result show that this resummation captures kinematic twists but not genuine higher-twist multiple-scattering corrections \cite{Boussarie:2021ybe}.

The remainder of this paper is organized as follows. In Sec.~\ref{sec:conventions}, we introduce the kinematics, conventions, and power counting used throughout the calculation. In Sec.~\ref{sec:prpgt}, we reorganize the background-field propagator by resumming the transverse kinematic corrections and isolating the interactions with the background field through an endpoint transformation, providing the basis for the simultaneous treatment of the eikonal and twist expansions. In Sec.~\ref{sec:dijet_amp}, we derive the DIS dijet production amplitudes for longitudinally and transversely polarized photons and study their eikonal and back-to-back limits. In Sec.~\ref{sec:itmd_rep}, we discuss the relation of our result to the iTMD framework. We summarize our results in Sec.~\ref{sec:summary}. Technical details are collected in the appendices.

\section{Theoretical setup}
\label{sec:conventions}

We consider the production of a quark--antiquark pair in DIS,
$\gamma^*(q)+A\to q(k_1)+\bar q(k_2)+X$, with the target represented by a background gluon field.  The virtual photon is chosen to carry a large minus momentum.  We use $v^\pm=(v^0\pm v^3)/\sqrt{2}$ and order components as $(+,-,\perp)$, so that
\begin{equation}
 v\cdot w=v^+w^-+v^-w^+-\vperp\cdot\wperp,
 \qquad
 g^{+-}=1,
 \quad
 g^{ij}=-\delta^{ij}.
\end{equation}
The photon momentum is fixed by $q^2=-Q^2$, $q^->0$, and $\qperp=0$, which gives $q^+=-Q^2/(2q^-)$.

For massless final-state quarks, minus-momentum conservation is implemented by
\begin{equation}
 z=\frac{k_1^-}{q^-},
 \qquad
 \bar z=\frac{k_2^-}{q^-},
 \qquad
 \epsilon_f^2=z\bar z Q^2.
\label{eq:fractions}
\end{equation}
with $\bar z=1-z$.  The scale $\epsilon_f$ is the standard photon wave-function combination that appears after Fourier transforming the hard splitting. We introduce the transverse momentum imbalance and the relative hard momentum, which will be useful when comparing the general amplitude with its leading-twist TMD limit:
\begin{equation}
 \bDeltaPerp=\koneperp+\ktwoperp,
 \qquad
 \Pperp=\bar z\,\koneperp-z\,\ktwoperp,
\label{eq:dijet-momenta}
\end{equation}
with inverse relations $\koneperp=\Pperp+z\bDeltaPerp$ and $\ktwoperp=-\Pperp+\bar z\bDeltaPerp$.  The back-to-back limit is then the expansion in $|\bDeltaPerp|/|\Pperp|$.

We now specify the background-field conventions used in the propagator expansion.  The background field is taken to be independent of $x^+$ at the
accuracy considered.  We absorb the gauge coupling into the background field and write $\Pop_\mu=\mathrm{p}_\mu+A_\mu$,\footnote{Momentum operators are set in roman face, for example $\mathrm{p}_\mu$ and $\mathrm{P}_\mu$.} $iD_\mu=i\partial_\mu+A_\mu$, and $[D_\mu,D_\nu]=-iF_{\mu\nu}$.
Accordingly, $A_\mu$ and $F_{\mu\nu}$ below denote $gA_\mu^a t^a$ and $gF_{\mu\nu}^a t^a$, respectively.

Gauge invariance is maintained through straight Wilson lines. A fundamental Wilson line along the $x^-$ direction is defined by
\begin{equation}
 \WL{a^-}{b^-}{\xperp}
 =\cP\exp\!\left[
   i\int_{b^-}^{a^-}\dd\xi^- A^+(\xi^-,\xperp)
 \right].
\label{eq:wilson-line}
\end{equation}
Straight transverse links at fixed $x^-$ are denoted by
$[\xperp,\yperp]_{x^-}$,
\begin{align}
[\xperp,\yperp]_{x^-}
=
\cP\exp\left[
i\int_0^1du\,
(x-y)^kA_k(x^-,{\mathbf \xi}_u)
\right],
\label{eq:trwilson-line}
\end{align}
where
$\xi_u^k=ux^k+(1-u)y^k$.
These links retain the gauge fields at light-cone infinity, so that no boundary condition in light-cone gauge is required.

Where it is convenient we use 
\begin{align}
    \bar F_{\mu \nu}(z^-, \zperp) = [\infty, z^-]_{\zperp} F_{\mu \nu}(z_-, \zperp)  [z^-, \infty]_{\zperp}\,.
    \label{eq:decF}
\end{align}

The calculation employs two complementary power-counting schemes. The eikonal expansion organizes corrections in inverse powers of the projectile energy, whereas the twist expansion includes both kinematic and dynamical twist contributions. We retain all terms required to reproduce either limiting description. Consequently, the final result contains all kinematic corrections associated with the leading-eikonal approximation together with all $x>0$ contributions required by the leading-twist TMD framework. Terms that are simultaneously sub-eikonal and beyond leading twist lie outside the stated accuracy and are discarded whenever such field-strength structures arise.

\section{Propagators resumming kinematic corrections}  \label{sec:prpgt}

In Ref.~\cite{Mukherjee:2026cte}, the scalar and quark propagators were expanded in both kinematic and dynamical twists.  Here we reorganize that derivation so that the kinematic corrections required by the eikonal limit are resummed while retaining all terms needed for the TMD limit.

The starting point for both the scalar and fermionic cases is the propagator in the mixed representation
\begin{align}
\label{eq:mixed_prop}
(k | \frac{1}{P^2 + \frac{\sigma F}{2} + i\epsilon }  | y)
= \int d^4 x  \,e^{i k \cdot x} \,\frac{1 }{2 k^-} (x| \frac{1}{{\rm P}^+ - \tfrac{\Ptop^2}{2k^-}+\frac{\sigma F}{4 k^-} + i\epsilon }| y)\,.
\end{align}
 Here $k^->0$, and we use the shorthand  $\sigma F\equiv\sigma^{\mu\nu}F_{\mu\nu}$, with $\sigma^{\mu\nu}\equiv \frac{i}{2}[\gamma^\mu,\gamma^\nu]$.  In the scalar case this spin-coupling term is set to zero.  We also take the gluon background to be independent of $x^+$, consistently with the stated twist accuracy.  The corrections associated with restoring the $x^+$ dependence were described in Ref.~\cite{Mukherjee:2026cte}.

We now analyze the matrix element on the right-hand side of Eq.~\eqref{eq:mixed_prop}.  It is useful to separate two aspects of the problem.  We first consider the scalar propagator, temporarily dropping the spin coupling $\sigma F$, in order to expose the transverse kinematic evolution and its endpoint transformation.   The fermionic propagator can then be obtained without repeating the scalar derivation. The mixed longitudinal-transverse spin interaction $\sigma^{-i}F_{-i}$ is absorbed by an exact fermionic endpoint rotation, while the transverse spin interaction $\sigma^{ij}F_{ij}$ enters through the replacement $\Ptop^2\to\Ptop^2-\sigma^{ij}F_{ij}/2$ in the transformed Hamiltonian.

\subsection{The scalar propagator}

In Ref.~\cite{Mukherjee:2026cte}, the scalar and quark propagators were expanded simultaneously in kinematic and dynamical twists. Here we reorganize that expansion into a form adapted to the interpolation considered in this work.  The transverse kinematic corrections are resummed to preserve the all-twist structure of the eikonal limit, while the non-zero-$x$ field-strength
insertions required by the leading-twist TMD limit are kept explicitly. We begin with the expansion introduced in Ref.~\cite{Mukherjee:2026cte}:
\begin{align}
\label{eq:expansionPperp}
  (x | \frac{1}{{\rm P}^+ - \tfrac{\Ptop^2}{2k^-} + i\epsilon }  | y)
 &=\,
 \delta(x^+ - y^+)\,
 \theta(x^- - y^-)\,(\xperp \,|\,
 \Big[
   -i\,[x^-,y^-] 
   - \int_{y^-}^{x^-} dz^- \,
     [x^-,z^-]\,
     \frac{\Ptop^2(z^-)}{2k^-}\,
     [z^-,y^-]
 \notag\\
 &
   + i \int_{y^-}^{x^-} dz_1^- \,
     [x^-,z_1^-]\,
     \frac{\Ptop^2(z_1^-)}{2k^-}
     \int_{y^-}^{z_1^-} dz_2^- \,
     [z_1^-,z_2^-]\,
     \frac{\Ptop^2(z_2^-)}{2k^-}\,
     [z_2^-,y^-]
   + \cdots
 \Big]|\, \yperp)
 \, ,
\end{align}
where we used the identity~\footnote{See e.g. Ref.~\cite{Mukherjee:2026cte} for the proof.}
\begin{align}
    (x^-| \frac{1}{\Pop^+ + i\varepsilon} |y^-) = - i \theta(x^--y^-) [x^-, y^-]\,.
\end{align}

Equation~\eqref{eq:expansionPperp} exhibits the characteristic Dyson-series structure generated by repeated insertions of the transverse kinetic operator between Wilson lines. Rather than treating these insertions order by order, it is advantageous to resum the entire series into a path-ordered exponential. This resummation treats the transverse propagation exactly while preserving the longitudinal gauge structure encoded in the Wilson lines. It also provides a natural starting point for separating purely kinematic propagation from genuine interactions with the background field. Accordingly, Eq.~\eqref{eq:expansionPperp} can be written in the compact form \footnote{A conceptually related technique, but with a different resummation, was used in Ref.~\cite{Kar:2026vzk}.}
\begin{align}
\label{eq:exponentiatedPperp}
 (x | \frac{1}{{\rm P}^+ - \tfrac{\Ptop^2}{2k^-} + i\epsilon }  | y)
 =&\,
 -i\delta(x^+ - y^+)\,
 \theta(x^- - y^-)\,
 (\xperp |
\mathcal U(x^-,y^-)
 |\yperp)
 \, ,
\end{align}
where
\begin{align}
\mathcal U(x^-,y^-)=
 \mathcal P
 \exp\Big(
 -i
 \int_{y^-}^{x^-}
 dz^-\,
 H(z^-;x^-)
 \Big)
 [x^-,y^-], \qquad H(z^-;x^-)
=
[x^-,z^-]
\frac{\Ptop^2(z^-)}{2k^-}
[z^-,x^-],
\end{align}
and $\mathcal P$ denotes ordering with respect to the longitudinal coordinate $z^-$. 

The path-ordered operator $\mathcal U(x^-,y^-)$ above makes it possible to regard the scalar propagator as an evolution along the light-cone coordinate $x^-$. In this language, the transverse kinetic operator plays the role of an effective Hamiltonian, while the Wilson line ensures gauge covariance. The evolution operator $\mathcal U(x^-,y^-)$ therefore satisfies the first-order equation
\begin{align}
\label{eq:DU(x,y)}
D_{x^-}\mathcal U(x^-,y^-)
=
-iH_0(x^-)\,
\mathcal U(x^-,y^-),
\end{align}
as $D_{x^-}[x^-,y^-]\,=\,0$ owing to the definition of the Wilson line.  
Here we introduced 
\begin{align}
H_{0}(x^-, k^-)  &= \frac{\Ptop^2(x^-)}{2k^-}\,.
\label{eq:H0}
\end{align}

Although Eq.~\eqref{eq:DU(x,y)} already resums the transverse propagation to all orders, the free transverse evolution remains part of the evolution kernel. For the interpolation developed in this work, it is advantageous to separate this purely kinematic propagation from the genuine interaction with the background field. This is accomplished by the endpoint transformation
\begin{align}
\mathcal U_s(x^-,y^-)
=
R^{-1}(x^-,k^-)\,
\mathcal U(x^-,y^-)\,
R(y^-,k^-),
\qquad
R(x^-,k^-)
=
e^{-ix^-\frac{\Ptop^2(x^-)}{2k^-}}\,,
\end{align}
which shifts the free transverse propagation into the endpoint factors.

The transformed evolution operator again satisfies a first-order evolution equation,
\begin{align}
D_{x^-}\mathcal U_s(x^-,y^-)
&=
D_{x^-}R^{-1}(x^-,k^-)\,
\mathcal U(x^-,y^-)\,
R(y^-,k^-)
+
R^{-1}(x^-,k^-)\,
D_{x^-}\mathcal U(x^-,y^-)\,
R(y^-,k^-)
\nonumber\\
&=
-i\,H_{0s}(x^-,k^-)\,
\mathcal U_s(x^-,y^-),
\end{align}
where the dependence on $k^-$ is temporarily suppressed. The transformed Hamiltonian is therefore
\begin{align}
H_{0s}(x^-,k^-)
=
H_0(x^-,k^-)
-
iR^{-1}(x^-,k^-)\,
D_{x^-}R(x^-,k^-)\,.
\end{align}
To determine $H_{0s}$ explicitly, we evaluate the covariant derivative of the endpoint rotation,
\begin{align}
D_{x^-}R(x^-,k^-)
\,&=\,-i\int_{0}^1\,d\alpha\,R^\alpha(x^-,k^-)\,D_-\Big(x^-\,\frac{\Ptop^2(x^-)}{2k^-}\Big)\,R^{1-\alpha}(x^-,k^-)\,
\na &  \,=\,-i R(x^-,k^-) H_{0}(x^-,k^-) -\frac{i\,x^-}{2k^-}\int_{0}^1\,d\alpha\,R^\alpha(x^-,k^-)\,\{{\mathrm P}_i, F_{-i}\}\,R^{1-\alpha}(x^-,k^-)\,.  
\end{align}
Substituting this result into the expression for $H_{0s}$ yields
\begin{align}
H_{0s}(x^-,k^-)
=
-
\frac{x^-}{2k^-}
\int_0^1 d\alpha\,
R^{-\alpha}(x^-,k^-)\,
\{{\mathrm P}_i(x^-),F_{-i}(x^-)\}\,
R^\alpha(x^-,k^-)\,,
\end{align}
where we have changed variables according to $\alpha\rightarrow1-\alpha$ and restored the explicit dependence on $k^-$.

Using the transformed evolution operator, the scalar propagator becomes
\begin{align}
 &(x | \frac{1}{{\rm P}^+ - \tfrac{\Ptop^2}{2k^-} + i\epsilon }  | y)
 =\,
 -i\delta(x^+ - y^+)\,
 \theta(x^- - y^-)\,(\xperp \,|R(x^-, k^-)\mathcal{U}_{s}(x^-,y^-)R^{-1}(y^-, k^-)|\, \yperp) \na &
 = -i\delta(x^+ - y^+)\,
 \theta(x^- - y^-)\,(\xperp \,|R(x^-, k^-)\Bigg[\mathcal{P}\exp\Big(-i \int_{y^-}^{x^-}\,dz^-\,H_s(z^-;x^-) \Big)[x^-,y^-]\Bigg]R^{-1}(y^-, k^-)|\, \yperp)\,,
\end{align}
where
\begin{align}
H_s(z^-;x^-)
&= [x^-,z^-]H_{0s}(z^-,k^-)[z^-,x^-]\na
&=
-
\frac{z^-}{2k^-}
\int_0^1 d\alpha\,
[x^-,z^-]\,
R^{-\alpha}(z^-,k^-)\,
\{{\mathrm P}_i(z^-),F_{-i}(z^-)\}\,
R^\alpha(z^-,k^-)\,
[z^-,x^-].
\end{align}
The endpoint rotation has therefore isolated the free transverse propagation into the endpoint factors $R$ and $R^{-1}$, while the evolution kernel itself is expressed entirely in terms of field-strength insertions. This representation provides the natural starting point for the interpolation developed in the following sections. In particular, it preserves the all-order transverse kinematics required in the eikonal limit while making the non-zero-$x$ interaction with the background field explicit.

\subsection{The quark propagator}

We now extend the construction of the previous subsection to the quark propagator.  The scalar evolution derived above carries over directly, except for the additional spin couplings to the background field.  In particular, the term involving $\sigma_{i-}F^{i-}$ requires a separate treatment before the result can be cast into the same evolution form as in the scalar case. The relevant propagator can be written as
\begin{align}
    (x|\frac{1}{\Pop^+ -\frac{\Ptop^2}{2k^-} + \frac{\sigma F}{4k^-} + i\epsilon} |y)\,=\,(x|\frac{1}{\Pop^+ - \frac{1}{2k^-}
 (
\Ptop^2-\frac{1}{2}\sigma^{ij}F_{ij}
 ) + \frac{\sigma^{-i} F_{-i}}{ 2 k^-} + i \epsilon}|y),
\end{align}  
Here the transverse spin coupling has been grouped with the transverse kinetic term, while the term proportional to \(\sigma^{-i}F_{-i}\) requires separate treatment.  The latter can be absorbed into fermionic endpoint rotations, leaving the remaining spin dependence to be incorporated into the longitudinal evolution.
\begin{align}
\Lambda_f\,&=\,1\,+\,i\frac{\sigma^{-l} \Pop_l}{2k^-} = \frac{\gamma^- \gamma^
+}{2} +\frac{ \slashed{\Pop} \gamma^-  }{2k^-}, \\
\Lambda_f^{-1}\,&= \,1\,-\,i\frac{\sigma^{-l} \Pop_l}{2k^-} =  \frac{\gamma^+ \gamma^
-}{2} + \frac{\gamma^-   \slashed{\Pop} }{2k^-},
\end{align}
where \(\Pop^-=k^-\). Note that $\Lambda_f$ is $P^+$ independent because it drops out in the combination $\gamma^-   \slashed{\Pop}$.  In terms of these rotations, the denominator satisfies the exact similarity transformation
\begin{align}
\label{eq:id_transverse}
\Pop^+ - \frac{1}{2k^-}
 (
\Ptop^2-\frac{1}{2}\sigma^{ij}F_{ij} )+ \frac{\sigma^{-i} F_{-i}}{2k^-}\,=\,{ \Lambda_f\, \bigg(\Pop^+ -  \frac{1}{2k^-}
 (
\Ptop^2-\frac{1}{2}\sigma^{ij}F_{ij} ) \bigg)\,\Lambda_f^{-1}}.
\end{align}
Consequently, the quark propagator can be expressed as
\begin{align}
 (x|\frac{1}{\Pop^+ -\frac{\Ptop^2}{2k^-} + \frac{\sigma F}{4k^-} + i\epsilon} |y)\,\,=\,(x|\,{ \Lambda_f\,\frac{1}{\Pop^+ -  \frac{1}{2k^-}
 (
\Ptop^2-\frac{1}{2}\sigma^{ij}F_{ij} )   + i \epsilon}\Lambda_f^{-1}} \,|y).
\end{align}
This representation isolates the \(F_{-i}\) spin coupling into the endpoint factors, while the remaining evolution retains the same structure as in the scalar case, with the replacement
\begin{equation}
\Ptop^2
\longrightarrow
\Ptop^2-\frac{1}{2}\sigma^{ij}F_{ij}.
\end{equation}
The construction of the previous subsection can therefore be applied directly, with the transverse spin interaction incorporated into the evolution Hamiltonian. One obtains
\begin{align}
 &(x|\frac{1}{\Pop^+ -\frac{\Ptop^2}{2k^-} + \frac{\sigma F}{4k^-} + i\epsilon} |y)\,\,=\, -i\delta(x^+ - y^+)\,
 \theta(x^- - y^-) \na & \times \,(\xperp \,|\Lambda_f(x^-,k^-)\,R(x^-, k^-)\Bigg[\mathcal{P}\exp\Big(-i \int_{y^-}^{x^-}\,dz^-\,H_{s\sigma}(z^-;x^-) \Big)[x^-,y^-]\Bigg]R^{-1}(y^-, k^-)\,\Lambda_f^{-1}(y^-,k^-)|\, \yperp)\,,
\end{align}
where the spin-dependent generalization of the scalar evolution Hamiltonian is
\begin{align}
H_{s\sigma}(z^-;x^-)\,&=\,H_{s}(z^-;x^-) - [x^-,z^-]R^{-1}(z^-, k^-)\frac{\sigma^{ij} F_{ij}(z^-)}{4k^-} R(z^-, k^-) [z^-,x^-]\, \na &
=\,-\,\frac{1}{2k^-}\int_{0}^1\,d\alpha\,[x^-,z^-]R^{-\alpha}(z^-, k^-)\,\bigg(z^-\,\{{\mathrm P}_i(z^-), F_{-i}(z^-)\}\,+\,\delta(1-\alpha)\,\frac{\sigma^{ij} F_{ij}}{2}\bigg)\,R^\alpha(z^-, k^-)[z^-,x^-]\,.
\end{align}
The first contribution is inherited from the scalar evolution and describes the coupling of transverse motion to the background field, whereas the second accounts for the spin interaction with the transverse field strength.  In this form, both contributions are embedded in the same gauge-covariant longitudinal evolution operator, while the remaining \(F_{-i}\) dependence is encoded in the endpoint rotations.

For the scattering amplitudes considered below, the relevant object is the propagator with an asymptotic on-shell quark leg.  Applying the LSZ reduction as in Refs.~\cite{Mukherjee:2026cte,Mukherjee:2026six}, the outgoing-quark propagator takes the form
\begin{align}
\label{eq:symmetric_prop}
&\lim_{k^2 \rightarrow 0}\,k^2\,(k | \frac{1}{\Pop^2 + \frac{\sigma F}{2} +i\epsilon }  | y)
 \na & =\,e^{ik^- y^+}(\kperp \,|\Lambda_f(\infty,k^-)\,\Bigg[\mathcal{P}\exp\Big(-i \int_{y^-}^{\infty}\,dz^-\,H_{s\sigma}(z^-;\infty;k^-) \Big)[\infty,y^-]\Bigg]R^{-1}(y^-,k^-)\Lambda_f^{-1}(y^-,k^-)|\, \yperp)\,.
\end{align}
It organizes the interaction of the outgoing quark with the background field into a longitudinal Wilson line dressed by transverse and spin-dependent corrections, with the fermionic endpoint factors kept explicitly.

The antiquark expression is obtained by reversing the momentum flow and the ordering of the background-field evolution.  For \(k^->0\), it reads
\begin{align}
&\lim_{k^2 \rightarrow 0}\,k^2\,(y | \frac{1}{\Pop^2 +\frac{\sigma F}{2}+ i\epsilon }  | -k)
 \na & =\,e^{ik^- y^+}(\yperp \,| \Lambda_f(y^-, -k^-)  R(y^-, -k^-)\Bigg[[y^-,\infty]\mathcal{P}\exp\Big(i \int_{y^-}^{\infty}\,dz^-\,H_{s\sigma}(z^-;\infty;-k^-) \Big)\Bigg] \Lambda_f^{-1}(\infty, -k^-)  |\, -\kperp)\,.
\end{align}
 Together, the quark and antiquark propagators provide the fermionic counterparts of the scalar result derived above and will serve as the building blocks for the background-field expansion of the dijet amplitude.

\section{Dijet production amplitudes} \label{sec:dijet_amp}

We now apply the background-field propagators derived in the previous section to DIS dijet production.  The amplitude for the splitting of a virtual photon with momentum \(q\) into an outgoing quark-antiquark pair with momenta \(k_1\) and \(k_2\) is
\begin{align}\label{eq:AmplitudeM}
  i \mathcal{M}=
  &\, i e \,\epsilon_\rho(q)
  \int d^4y \, e^{-iq\cdot y}
  \lim_{k_{1}^2, k_2^2 \to 0}
  \bar{u}(k_1)\,k_1^2\,
 (k_1|
  \frac{1}{{\rm P}^2+\frac{1}{2}\sigma F}
  |y )
  \gamma^\rho
  (y|
  \frac{1}{{\rm P}^2+\frac{1}{2}\sigma F} 
 |-k_2)\,
  k_2^2\,v(k_2)\,.
\end{align}
The two propagators describe the subsequent interaction of the produced quark and antiquark with the target background field.  Their resummed form derived above allows these interactions to be organized in terms of longitudinal Wilson lines dressed by transverse-motion and spin-dependent corrections.

The DIS dijet amplitude through twist-three accuracy was derived in Ref.~\cite{Mukherjee:2026cte} by expanding systematically in the inverse hard momentum.  Here we retain the resummed propagators of Eq.~\eqref{eq:symmetric_prop} instead.  This representation keeps the kinematic dependence associated with transverse propagation unexpanded and provides a convenient starting point for studying both the leading-twist limit and the high-energy, small-\(x\) limit within the same expression.  We consider the longitudinal and transverse photon polarizations separately.

\subsection{Longitudinal photon polarization}

For a longitudinally polarized photon, we take
\begin{equation}
\epsilon_\rho^{,L}(q)
= \frac{Q}{q^-}\,g_{+\rho},
\end{equation}
so that the photon vertex is proportional to \(\gamma^-\).  A useful simplification then occurs: the fermionic endpoint rotations adjacent to the photon vertex are projected out by the light-cone Dirac structure.  Using \(\Pop^-=k^-\), one finds
\begin{align}
    \Lambda_f^{-1} \gamma^-  = 
    \Lambda_f \gamma^-  =  \gamma^-,
\end{align}
The LSZ-reduced quark propagator therefore simplifies to
\begin{align}
\label{eq:symmetric_prop_long}
&\lim_{k^2 \rightarrow 0}\,k^2\,(k | \frac{1}{\Pop^2 + \frac{\sigma F}{2} +i\epsilon }  | y) \gamma^- 
 =  \,e^{ik^- y^+}(\kperp \,|\,\Bigg[\mathcal{P}\exp\Big(-i \int_{y^-}^{\infty}\,dz^-\,H_{s\sigma}(z^-;\infty;k^-) \Big)[\infty,y^-]\Bigg]R^{-1}(y^-,k^-)|\, \yperp)\,   {\gamma^-}
\end{align}
Thus, for longitudinal polarization, the spin dependence associated with the endpoint rotations drops out explicitly, while the spin-dependent interaction contained in \(H_{s\sigma}\) remains part of the longitudinal evolution.

Substituting the corresponding quark and antiquark propagators into Eq.~\eqref{eq:AmplitudeM}, we obtain
\begin{align}\label{eq:Amp_L}
  i \mathcal{M}_L=
  &\, i e \,\epsilon_-(q)
  \int d^4y \, e^{-iq\cdot y}\,e^{i (k_1^- + k_2^-)y^+}\,
  \bar{u}(k_1)\,(\koneperp \,|\Bigg[\mathcal{P}\exp\Big(-i \int_{y^-}^{\infty}\,dz^-\,H_{s\sigma}(z^-;\infty;k_1^-) \Big)[\infty,y^-]\Bigg]R^{-1}(y^-,k_1^-)|\, \yperp)\na &
  \gamma^-
  (\yperp \,|R(y^-, -k_2^-)\Bigg[[y^-,\infty]\mathcal{P}\exp\Big(i \int_{y^-}^{\infty}\,dz^-\,H_{s\sigma}(z^-;\infty;-k_2^-) \Big)\Bigg]|\, -\ktwoperp)\, 
 v(k_2)\,.
\end{align}
This result provides a resummed representation of the longitudinal dijet amplitude.  To organize the subsequent analysis, we separate the contribution obtained by setting both \(H_{s\sigma}\) evolution operators to unity from those involving explicit insertions of \(H_{s\sigma}\).  We refer to the former as the semi-infinite dipole contribution and consider it first.

\subsubsection{Semi-infinite dipole contribution}
We first consider the contribution obtained by setting the two
\(H_{s\sigma}\) evolution operators in Eq.~\eqref{eq:Amp_L} to unity.  The semi-infinite dipole contribution to the 
longitudinal dijet amplitude is
\begin{align}\label{eq:AmplitudeM0}
  i \mathcal{M}_{0,L}
 =
  &\, i e \,\epsilon_-(q) \bar{u}(k_1)   \gamma^-\, v(k_2)\,
  \int d^4y\,\int d^2 \zoneperp d^2 \ztwoperp \, e^{-iq\cdot y}\,e^{i (k_1^- + k_2^-)y^+}\,e^{-i \koneperp\cdot\zoneperp -i \ktwoperp\cdot\ztwoperp}\,
  \na & \times 
  \,[\infty,y^-]_{\zoneperp}
  (\zoneperp|
  R^{-1}\left(y^-, \kappa^- \right)
  \,|\ztwoperp)\,[y^-,\infty]_{\ztwoperp}\,,
\end{align}
where
\begin{align}
    \kappa^-  =  \left(\frac{1}{k_1^-} + \frac{1}{k_2^-} \right)^{-1}.
\end{align}
\begin{figure}[t]
    \begin{subfigure}[t]{0.32\linewidth}
        \centering
        \includegraphics[width=\linewidth]{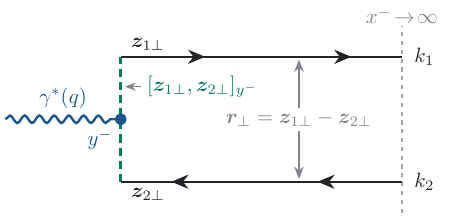}
        \caption{}
        \label{fig:sym_con}
    \end{subfigure}
    \hfill
    \begin{subfigure}[t]{0.32\linewidth}
        \centering
        \includegraphics[width=\linewidth]{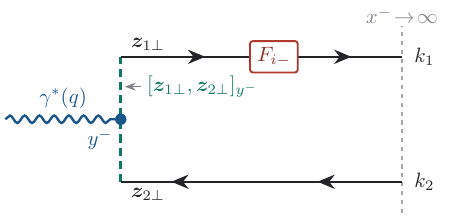}
        \caption{}
        \label{fig:q_con}
    \end{subfigure}
    \hfill
    \begin{subfigure}[t]{0.32\linewidth}
        \centering
        \includegraphics[width=\linewidth]{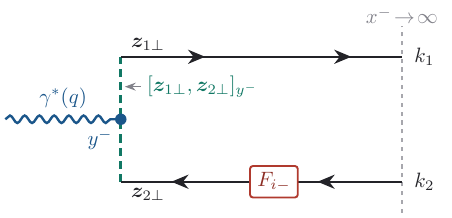}
        \caption{}
        \label{fig:aq_con}
    \end{subfigure}
    \caption{%
    \justifying
    Operator structures contributing to the longitudinal dijet amplitude.
    (a) Semi-infinite dipole contribution, Eq.~\eqref{eq:AmplitudeM0}.
    The two horizontal lines are the longitudinal Wilson lines
    $[\infty,y^-]_{\zoneperp}$ and $[y^-,\infty]_{\ztwoperp}$,
    which extend to $x^-\to\infty$ at transverse positions
    $\zoneperp$ and $\ztwoperp$ and carry the outgoing momenta
    $k_1$ and $k_2$. The vertical dashed segment is the transverse
    gauge link $[\zoneperp,\ztwoperp]_{y^-}$ contained in the
    transverse evolution operator $R^{-1}(y^-,\kappa^-)$.
    {(b) Semi-infinite dipole contribution with a field strength insertion in the quark longitudinal Wilson line $[y^-,\infty]$.
    (c) Semi-infinite dipole contribution with a field strength insertion in the antiquark longitudinal Wilson line $[\infty,y^-]$}.
    }
    \label{fig:operator_basis}
\end{figure}

The corresponding operator structure is shown in Fig.~\ref{fig:sym_con}.

For this contribution, the transverse evolution operator \(R^{-1}\) can be expanded to the accuracy relevant here.  As shown in Appendix~\ref{sec:TEexpansion},
\begin{align}
(\zoneperp |   R^{-1}\left(y^-, \kappa^- \right)    |\ztwoperp )\,\approx\,(\zoneperp| e^{i  \frac{y^-\,p_{\perp}^2}{2 \kappa^-}  }|\ztwoperp)\,[\zoneperp, \ztwoperp]_{y^-},
\end{align}
The correction proportional to \(y^-F_{ij}\) is sub-sub-eikonal and lies
beyond the twist accuracy considered here.
Here and below, we use
  \begin{align}
      \bar \delta(x) \equiv 2\pi \, \delta(x) ,
      \qquad
      \int \bar d x \equiv \int\frac{dx}{2\pi}.
  \end{align}
The free transverse kernel can then be written in momentum space as
\begin{align}
    \label{eq:z1_Free_Prop_z2}
    (\zoneperp|e^{i \frac{y^-\,p_{\perp}^2}{2 \kappa^-}  } \,|\ztwoperp) 
     = \int { \bar d}^2 \pperp \;
     e^{i (\zoneperp - \ztwoperp) \cdot\pperp  }  e^{i \frac{y^-\,p_{\perp}^2}{2 \kappa^-}  }\,.
\end{align}
Substituting this representation into the amplitude and performing the
\(y^+\) and transverse-coordinate integrations associated with the photon
momentum, we obtain
\begin{align}\label{eq:AmplitudeM2}
  i \mathcal{M}_{0,L}
 & = i e \,\epsilon_-(q) \bar{u}(k_1)\,\gamma^-\,v(k_2)\, { {\bar\delta(k_g^-) }} \,  \int d^2 \zoneperp  d^2 \ztwoperp \,e^{-i \koneperp\cdot\zoneperp -i \ktwoperp\cdot\ztwoperp}\,
  \int { \bar d}^2 \pperp \;
     e^{i (\zoneperp - \ztwoperp) \cdot\pperp  }
  \int dy^-  \,\, e^{i k_{p_\perp}^+   y^-}
  \na &
  [\infty,y^-]_{\zoneperp} [\zoneperp,\ztwoperp]_{y^-} 
 \,[y^-,\infty]_{\ztwoperp}\,,
\end{align}
where $k_g^- \equiv k_1^-+k_2^- - q^-$ is the minus component of the gluon momentum exchanged with the background field. We have also accounted for $\qperp=0$ and defined
\begin{align}
\label{Eq:k^+}
k^+_{p_\perp}
=
\frac{p_\perp^2}{2\kappa^-}-q^+
=
\frac{p_\perp^2+\epsilon_f^2}{2q^- z\bar z}
>0\,.
\end{align}
The remaining transverse-momentum integral is evaluated explicitly in Eq.~\eqref{eq:p-integral}.
\subsubsection{{Contributions with explicit field strength insertions}}

The contribution \(i\mathcal M_{0,L}\) obtained above does not by itself reproduce the complete amplitude in the kinematic limits of interest. We therefore restore the terms generated by the ordered evolution operator {in Eq.~\eqref{eq:Amp_L}}. As we show later, these contributions are required to reproduce the fixed-twist expansion in the back-to-back limit. We should note that the eikonal $x\rightarrow 0$ dijet amplitude can be entirely reproduced by the semi-infinite dipole contribution. We first consider the insertion on the quark line and then the corresponding antiquark contribution.

Expanding the quark-line evolution operator in Eq.~\eqref{eq:Amp_L} to the required order gives
\begin{align}
i \mathcal{M}_{q,L}=
  &\, i e \,\epsilon_-(q)
  \int d^4y \, e^{-iq\cdot y}\,e^{i (k_1^- + k_2^-)y^+}\,
  \bar{u}(k_1)\,(\koneperp \,|i\frac{1}{2k_1^-}\int_{y^-}^{\infty}\,dz^-\,\int_{0}^1\,d\alpha\,[\infty,z^-]R^{-\alpha}(z^-,k_1^-)\,\Big(z^-\{{\mathrm P}_i(z^-), F_{-i}(z^-)\}\na & +\,\delta(1-\alpha)\,\frac{\sigma^{ij} F_{ij}(z^-)}{2}\Big)\,R^\alpha(z^-,k_1^-)[z^-,y^-]R^{-1}(y^-,k_1^-)|\, \yperp)
  \gamma^-
  (\yperp \,|R(y^-, -k_2^-)[y^-,\infty]|\, -\ktwoperp)\, 
 v(k_2)\,.
\end{align}

The $z^-$-dependent transverse rotations can be commuted through the operator insertion and moved onto the external transverse-momentum states.  The commutators generated in this step are simultaneously sub-sub-eikonal and of higher twist and therefore lie beyond the accuracy considered here.  The rotations acting on the external states then reduce to a transverse kinetic phase.  Combining the remaining rotations at $y^-$ gives
\begin{align}
i \mathcal{M}_{q,L}=
  &\, i e \,\epsilon_-(q)
  \int dy^- dy^+  \, e^{-iq^- y^+ -iq^+ y^- }\,e^{i (k_1^- + k_2^-)y^+}\,
  \frac{i}{2k_1^-}\int_{y^-}^{\infty}\,dz^-\,\int_{0}^1\,d\alpha\,
  e^{i \alpha z^- \frac{k_{1\perp}^2 - k_{2\perp}^2}{2k_1^-} } 
  \bar{u}(k_1)\,(\koneperp \,|
  [\infty,z^-]\,\na & 
  \times \Big(z^-\{{\mathrm P}_i(z^-), F_{-i}(z^-)\} +\,\delta(1-\alpha)\,\frac{\sigma^{ij} F_{ij}(z^-)}{2}\Big)\,[z^-,y^-] 
  R^{-1}\left(y^-, \kappa^- \right) 
  [y^-,\infty] |\, -\ktwoperp)\, 
 \gamma^- v(k_2)\,.
 \end{align}

The spin-dependent insertion proportional to $\sigma^{ij}F_{ij}$ is suppressed in both power counting schemes considered here and will therefore be omitted.  The remaining $\alpha$ integration can be carried out explicitly.  Using the transverse matrix element derived in Appendix~\ref{app:MatrixElemPF}, the quark-line contribution becomes
\begin{align}
i \mathcal{M}_{q,L}=
  &\, i e \,\epsilon_-(q)   \bar{u}(k_1) \gamma^- v(k_2)\,
  \int dy^- dy^+  \, e^{-iq^- y^+ -iq^+ y^- }\,e^{i (k_1^- + k_2^-)y^+}\,
  \frac{1}{k_{1\perp}^2 - k_{2\perp}^2} \int d^2 \zoneperp d^2 \ztwoperp  
  e^{ - i \koneperp\cdot\zoneperp  - i \ktwoperp\cdot\ztwoperp }
  \na & 
  \!\int_{y^-}^{\infty}dz^-
\! \bigg(\! e^{i  z^- \frac{k_{1\perp}^2 \!- k_{2\perp}^2}{2k_1^-} } \!-\!1 \!\bigg)\!
  \int { \bar d}^2 \pperp 
     e^{i (\zoneperp \!-\! \ztwoperp) \cdot\pperp  }  e^{i \frac{y^- p_{\perp}^2}{\kappa^-}  }  
  (k_1 \!-\! k_2)_i 
  \bar F_{-i}(z^-, \zoneperp ) [\infty, y^-]_{\zoneperp} [\zoneperp, \ztwoperp]_{y^-} [y^-,\infty]_{\ztwoperp}\,.
\end{align}
The remaining transverse-momentum integral is Gaussian and can be evaluated explicitly:
\begin{equation}
\label{eq:p-integral}
{
\int \frac{d^2\pperp}{(2\pi)^2}\,
e^{\,i(\zoneperp-\ztwoperp)
\cdot\pperp}\,
e^{\,i\frac{y^-\, p_{\perp}^2}{\kappa^-}}
=
\frac{i\,k_1^-k_2^-}
{2\pi y^-(k_1^-+k_2^-)}
\exp\left[
-\frac{i\,k_1^-k_2^-}
{2y^-(k_1^-+k_2^-)}
(\zoneperp-\ztwoperp)^2
\right]
}
.
\end{equation}
After performing the $y^+$ integration and introducing $z=k_1^-/q^-$ and $\bar z=k_2^-/q^-$, we obtain
\begin{align}\label{eq: quark_l_amp}
i \mathcal{M}_{q,L}=
  &\, - e \,\epsilon_-(q)   \bar{u}(k_1) \gamma^- v(k_2)\,
  z \bar z \,  { {q^- \, \delta(k_g^-) }}
  \int dy^- 
  \frac{e^{-iq^+ y^- }}{y^-(k_{1\perp}^2 - k_{2\perp}^2)} \int d^2 \zoneperp d^2 \ztwoperp  e^{ - i \koneperp\cdot\zoneperp  - i \ktwoperp\cdot\ztwoperp } 
  \na &
  \int_{y^-}^{\infty}dz^-
  \bigg( e^{i  z^- \frac{k_{1\perp}^2  - k_{2\perp}^2}{2k_1^-} } \!-\! 1 \bigg) 
  e^{ -\frac{i\,z \bar z q^-} {2y^-} ({z}_{1\perp}-{z}_{2\perp})^2 }
  (k_1 - k_2)_i \; 
  \bar F_{-i}(z^-, \zoneperp ) [\infty, y^-]_{\zoneperp} [\zoneperp, \ztwoperp]_{y^-} [y^-,\infty]_{\ztwoperp}\,.
\end{align}
{This amplitude is illustrated  in Fig.~\ref{fig:q_con}}. 

The insertion on the antiquark line is obtained in the same way.  It gives
\begin{align}
\label{eq: antiquark_l_amp}
i \mathcal{M}_{\bar{q},L}=
  &\, e \,\epsilon_-(q)   \bar{u}(k_1) \gamma^- v(k_2)\,
  z \bar z \, { {q^- \, \delta(k_g^-) }}
  \int dy^- 
  \frac{e^{-iq^+ y^- }}{y^-(k_{1\perp}^2 - k_{2\perp}^2)}\int d^2 \zoneperp d^2 \ztwoperp e^{ - i \ktwoperp\cdot\ztwoperp  - i \koneperp\cdot\zoneperp }
  \na &
  \int_{y^-}^{\infty}dz^-
  \bigg( e^{i  z^- \frac{k_{2\perp}^2  - k_{1\perp}^2}{2k_2^-} } \!-\! 1 \bigg) 
  e^{ -\frac{i\,z \bar z q^-} {2y^-} ({z}_{1\perp}-{z}_{2\perp})^2 }
  (k_1 - k_2)_i \; 
 [\infty,y^-]_{\zoneperp}[\zoneperp, \ztwoperp]_{y^-} [y^-, \infty]_{\ztwoperp} \bar F_{-i}(z^-, \ztwoperp )\,,
\end{align}
{see Fig.~\ref{fig:aq_con}}. Combining the contribution obtained in the preceding subsection with the single-insertion terms on the quark and antiquark lines, the complete longitudinal amplitude at the accuracy considered here can be written as
\begin{align}
\label{eq:Full_amplitude}
i\mathcal{M}_L\,=\,i\mathcal{M}_{0,L}\,+\,i\mathcal{M}_{q,L}\,+\,i\mathcal{M}_{\bar q,L}\,.
\end{align}
This form keeps the transverse propagation resummed while retaining the sub-eikonal operator insertions required at the present accuracy.  In the following, we examine how the resulting expression reduces in the relevant kinematic limits.

\subsubsection{Eikonal limit}

We first consider the high-energy limit of the resummed longitudinal amplitude.  This provides an important consistency check of the construction, since the finite-energy transverse propagation retained in the preceding derivation must reduce, in the appropriate limit, to the standard eikonal propagation through the background field.  In this regime the interaction of the energetic quark and antiquark with the target is encoded in lightlike Wilson lines, and the longitudinal photon amplitude should therefore reproduce the familiar dipole structure of the shock-wave formalism.

The relevant simplification can be seen directly from the contribution
$\mathcal M_{0,L}$.  Starting from Eq.~\eqref{eq:AmplitudeM2}, it is convenient to collect the Wilson lines connecting the quark and antiquark trajectories into the staple
\begin{align}
S(y^-)\,=\,[\infty,y^-]_{\zoneperp} [\zoneperp,\ztwoperp]_{y^-} 
  \,[y^-,\infty]_{\ztwoperp}\,.
\end{align}
The dependence on the longitudinal interaction point then enters through the Fourier integral
\begin{align}
\mathcal{I}\,=\,\int_{-\infty}^{\infty}\,dy^-\,e^{i k_{p_\perp}^+ y^-}\,S(y^-)\,.
\end{align}

{A direct expansion of the staple is not appropriate in the eikonal limit: the Wilson lines resum arbitrarily many interactions with the enhanced background field and must be kept to all orders.  Instead, the high-energy expansion can be organized by first rewriting the integral in terms of the longitudinal derivative of the staple,}
\begin{align}
\mathcal{I}\,=\,-\int_{-\infty}^{\infty}\,dy^-\,e^{i k_{p_\perp}^+ y^-}\,\Big(\int_{y^-}^{\infty} dz^- \frac{dS(z^-)}{dz^-} - S(\infty)\Big).
\end{align}
The boundary term is proportional to $\delta(k_{p_\perp}^+)$ and does not contribute for $k_{p_\perp}^+>0$.  Interchanging the order of integration therefore gives
\begin{align}
\mathcal{I}\,=\,-\int_{-\infty}^{\infty}\,dz^-\,\int_{-\infty}^{z^-}\,e^{i k_{p_\perp}^+ y^-}\,dy^- \frac{dS(z^-)}{dz^-}\,=\,\frac{i}{k_{p_\perp}^+}\,\int_{-\infty}^{\infty}\,dz^-\,e^{i k_{p_\perp}^+ z^-}\,\frac{dS(z^-)}{dz^-}.
\end{align}

The derivative of the staple generates a field-strength insertion $F_{i-}$ at $z^-$.  We then expand the remaining phase in the eikonal limit.  Keeping the leading term gives
\begin{align}
\mathcal{I}^{\,\mathrm{eik}}\,=\,\frac{i}{k_{p_\perp}^+}\,\int_{-\infty}^{\infty}\,dz^-\,\frac{dS(z^-)}{dz^-} = \frac{i}{k_{p_\perp}^+} \left( S(\infty) - S(-\infty) 
\right)
= \frac{i}{k_{p_\perp}^+} \left( 1 - [\infty,-\infty]_{\zoneperp} [\zoneperp,\ztwoperp]_{-\infty} 
  \,[-\infty,\infty]_{\ztwoperp} 
\right).
\end{align}
Thus the non-zero-$x$ staple appearing in the resummed amplitude continuously reduces to the standard eikonal Wilson-line structure.  The two terms have the usual interpretation: the unit operator corresponds to propagation without interaction, while the second term describes the eikonal scattering of the $q\bar q$ pair from the target.
Substituting this result into Eq.~\eqref{eq:AmplitudeM2}, we find
\begin{align}\label{eq:AmplitudeM0_eik_sub}
  i \mathcal{M}^{\rm eik}_{L}
 & = - 2 q^- z \bar z e \,\epsilon_-(q) \bar{u}(k_1)\,\gamma^-\,v(k_2)\, \bar \delta(k_g^-) \int d^2 \zoneperp  d^2 \ztwoperp \,e^{-i \koneperp\cdot\zoneperp -i \ktwoperp\cdot\ztwoperp}\,
  \int { \bar d}^2 \pperp \;
      \frac{e^{i (\zoneperp - \ztwoperp) \cdot\pperp  } }{p^2 + \epsilon_f^2}
  \na &
  \times \left( 1 - [\infty,-\infty]_{\zoneperp} [\zoneperp,\ztwoperp]_{-\infty} 
  \,[-\infty,\infty]_{\ztwoperp} 
\right)  \na 
 & = - 2\,  z \bar z{ { q^- }} e \,\epsilon_-(q) \bar{u}(k_1)\,\gamma^-\,v(k_2)\, {  { \delta(k_g^-) }} \int d^2 \zoneperp  d^2 \ztwoperp \,e^{-i \koneperp\cdot\zoneperp -i \ktwoperp\cdot\ztwoperp}\,
  K_0 (\epsilon_f |\zoneperp - \ztwoperp|)
  \na &
  \times \left( 1 - [\infty,-\infty]_{\zoneperp} [\zoneperp,\ztwoperp]_{-\infty} 
  \,[-\infty,\infty]_{\ztwoperp} 
\right).
 \end{align}
The appearance of the modified Bessel function $K_0$ is the standard transverse-coordinate representation of the longitudinal photon wave function, whereas the Wilson-line combination contains the complete eikonal interaction with the target.  Equation~\eqref{eq:AmplitudeM0_eik_sub} therefore has precisely the expected separation between the perturbative $\gamma_L^*\to q\bar q$ splitting and the eikonal scattering factor.

For longitudinal photon polarization, Eq.~\eqref{eq:AmplitudeM0_eik_sub} exhausts the leading-eikonal contribution.  It agrees with the high-energy result of Eq.~(187) of Ref.~\cite{Kar:2026vzk}.  If the transverse background field is further set to zero, the transverse link becomes trivial and the staple reduces to the conventional product of longitudinal Wilson lines, reproducing the standard CGC dijet amplitude of Refs.~\cite{Dominguez:2011wm,Dominguez:2011br}. The general resummed expression thus contains the usual small-$x$ result as its leading high-energy limit. We see that the contributions from Eq.~\eqref{eq: quark_l_amp} and Eq.~\eqref{eq: antiquark_l_amp}, see corresponding  Fig.~\ref{fig:q_con} and Fig.~\ref{fig:aq_con},  explicitly vanish when the phases in exponentials  are neglected in the strict eikonal limit. Therefore, these terms only become relevant when sub-eikonal corrections need to be reproduced.

\subsubsection{TMD limit}

We next consider the opposite kinematic expansion relevant for nearly back-to-back dijets.  In this regime the relative transverse momentum $\Pperp$ is hard, whereas the total transverse momentum $\bDeltaPerp=\koneperp+\ktwoperp$ remains parametrically smaller.  In coordinate space, this hierarchy corresponds to a small transverse separation between the quark and antiquark. The TMD limit can therefore be obtained by expanding the resummed amplitude in
\begin{equation}
\rperp=\zoneperp-\ztwoperp
\end{equation}
around the common impact parameter
\begin{equation}
\bperp=z\zoneperp+\bar z\ztwoperp,
\end{equation}
or equivalently,
\begin{equation}
\zoneperp=\bperp+\bar z\,\rperp,
\qquad
\ztwoperp=\bperp-z\,\rperp.
\end{equation}
At leading power in the back-to-back expansion, only the term linear in $\rperp$ contributes to the gluon TMD operator.
\begin{figure}[t]
    \centering
    \includegraphics[width=0.5\linewidth]{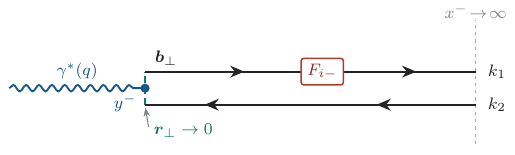}
    \caption{
    \justifying
    Leading-twist TMD limit of Fig.~\ref{fig:operator_basis}.  For nearly
    back-to-back jets the transverse separation of the pair is small,
    $\rperp\to0$, so that both Wilson lines sit at the common impact parameter $\bperp$.  At first order in $\rperp$ the staple of Fig.~\ref{fig:operator_basis} collapses onto a single field-strength insertion $F_{i-}$, evaluated at $(\xi^-,\bperp)$ and integrated over $\xi^-$ from $y^-$ to infinity.}
    \label{fig:tmd_con}
\end{figure}

The required expansion of the staple follows from the Mandelstam derivative of a longitudinal Wilson line,
\begin{align}
    \frac{1}{i}
    \bigl(\partial_i[\infty,y^-]_{\xperp}\bigr)
    [y^-,\infty]_{\xperp}
     &=
    A_i(\infty,\xperp)
    -[\infty,y^-]_{\xperp}
       A_i(y^-,\xperp)
       [y^-,\infty]_{\xperp}
    \notag\\
    &
    +\int_{y^-}^{\infty}d\xi^-\,
       [\infty,\xi^-]_{\xperp}
       F_{i-}(\xi^-,\xperp)
       [\xi^-,\infty]_{\xperp}.
\end{align}
At the same order in $\rperp$, the transverse Wilson line $[\zoneperp,\ztwoperp]{y^-}$ must also be expanded. Its contribution cancels the $A_i(y^-)$ endpoint term in the derivative above. Taking $A_i(\infty,\xperp)=0$, the first-order expansion of the complete staple is therefore
\begin{equation}
\begin{aligned}
    & [\infty,y^-]_{\zoneperp}
      [\zoneperp,\ztwoperp]_{y^-}
      [y^-,\infty]_{\ztwoperp}
 = 1
    +i\,r_{\perp}^{i}
      \int_{y^-}^{\infty}d\xi^-\,
      [\infty,\xi^-]_{\bperp}
      F_{i-}(\xi^-,\bperp)
      [\xi^-,\infty]_{\bperp}
    +\mathcal O(r_{\perp}^{2}).
\end{aligned}
\end{equation}
At this order the result is independent of the parameter $z$ specifying the position of the expansion point along the transverse segment.  This is also transparent diagrammatically: in the correlation limit the finite transverse separation in Fig.~\ref{fig:sym_con} collapses onto the single gluon insertion shown in Fig.~\ref{fig:tmd_con}.

We first apply this expansion to $\mathcal M_{0,L}$.  Keeping terms through first order in $\rperp$ gives
\begin{align}
  i \mathcal{M}_{0,L}^{\rm TMD}
 &\approx i e \,\epsilon_-(q) \bar{u}(k_1)\,\gamma^-\,v(k_2)\, \int d^2 \rperp  d^2 \bperp \,e^{-i \Pperp\cdot\rperp - i \bDeltaPerp\cdot\bperp }\,
  \int { \bar d}^2 \pperp \;
     e^{i \rperp \cdot\pperp  }
  \int dy^- dy^+ \,\, e^{i k_{p_\perp}^+   y^-}\,e^{- i k_g^- y^+}
  \na &
  \times \left( 
   1
    +i\,r_{\perp}^{i}
      \int_{y^-}^{\infty}d\xi^-\,
      [\infty,\xi^-]_{\bperp}
      F_{i-}(\xi^-,\bperp)
      [\xi^-,\infty]_{\bperp}
  \right)
 \,.
\end{align}
The unit term is proportional to $\delta(k_{p_\perp}^+)$ and vanishes for the physical kinematics $k_{p_\perp}^+>0$.  The first nonvanishing contribution is therefore linear in the transverse separation and directly generates the gluon field-strength operator characteristic of TMD factorization.  Performing the $y^-$ integration gives
\begin{align}
  i \mathcal{M}_{0,L}^{\rm TMD}
 &\approx 2\pi \delta (k_g^-) i e \,\epsilon_-(q) \bar{u}(k_1)\,\gamma^-\,v(k_2)\, \int d^2 \rperp  d^2 \bperp \,e^{-i \Pperp\cdot\rperp - i \bDeltaPerp\cdot\bperp }\,
      \int d\xi^-\,
  \int { \bar d}^2 \pperp \;
     e^{i \rperp \cdot\pperp  }
   \, \frac{e^{i k_{p_\perp}^+   \xi^-}}{k_{p_\perp}^+} 
  \na &
    \,r_{\perp}^{i}
      [\infty,\xi^-]_{\bperp}
      F_{i-}(\xi^-,\bperp)
      [\xi^-,\infty]_{\bperp}
 \,.
\end{align}
The $\rperp$ integration converts the transverse separation into a derivative with respect to the hard relative momentum,
\begin{align}
  i \mathcal{M}_{0,L}^{\rm TMD}
 &\approx 2\pi \delta (k_g^-) e \,\epsilon_-(q) \bar{u}(k_1)\,\gamma^-\,v(k_2)\, 
      \int d\xi^-\,
   \frac{\partial}{\partial P_i} \left( \frac{e^{i k_{P_\perp}^+   \xi^-}}{k_{P_\perp}^+} \right) 
 \int d^2 \bperp \,e^{ - i \bDeltaPerp\cdot\bperp }\,
      [\infty,\xi^-]_{\bperp}
      F_{i-}(\xi^-,\bperp)
      [\xi^-,\infty]_{\bperp}
 \,.
\end{align}
Using Eq.~\eqref{Eq:k^+}, this becomes
\begin{align}
  i \mathcal{M}_{0,L}^{\rm TMD}
 &\approx - 8\pi  e \,\epsilon_-(q) \bar{u}(k_1)\,\gamma^-\,v(k_2)\,  {  { \, \delta(k_g^-) }}    \frac{ z \bar z{  {\,q^-  }} }{(P_\perp^2 + \epsilon_f^2)^2}  
     \int d\xi^-\, e^{i k_{P_\perp}^+ \xi^-}
   \left( 1 - i { k_{P_\perp}^+ \xi^- } \right)  \na & \times
 \int d^2 \bperp \,e^{ - i \bDeltaPerp\cdot\bperp }\,
      [\infty,\xi^-]_{\bperp}
      P^i F_{i-}(\xi^-,\bperp)
      [\xi^-,\infty]_{\bperp}
 \,.
\end{align}

The above expression has an interesting form.  The first term in parentheses has the hard coefficient and the operator expected for the leading-twist gluon TMD amplitude.  The second term, however, originates from differentiating the non-zero-$x$ longitudinal phase and contains an additional factor $-ik_{P_\perp}^+\xi^-$.  It is not a part of the conventional leading-twist TMD hard factor.  Since the full resummed amplitude also contains the contributions $\mathcal M_{q,L}$ and $\mathcal M_{\bar{q},L}$, consistency requires this additional term to be canceled upon their back-to-back expansion.  We now demonstrate this cancellation explicitly.
For the quark-line contribution, we start from the general result in Eq.~\eqref{eq: quark_l_amp}.  At leading order in the correlation limit, the staple reduces to unity.  The $\ztwoperp$ integration can then be performed analytically, yielding
\begin{align}
i \mathcal{M}_{q,L}^{\rm TMD} &=
  \, - e \,\epsilon_-(q)   \bar{u}(k_1) \gamma^- v(k_2)\,
  z \bar z \, {  {q^- \, \delta(k_g^-) }}
  \int dy^- 
  \frac{e^{-iq^+ y^- }}{k_{1\perp}^2 - k_{2\perp}^2}\int_{y^-}^{\infty}\,dz^-\,
  \bigg( e^{i  z^- \frac{k_{1\perp}^2 - k_{2\perp}^2}{2k_1^-} } -1 \bigg)  
  \na &
\times \left\{-\frac{2\pi i}{z\bar{z}\,q^-}
e^{ 
\frac{i y^- \,k_{2\perp}^{\,2}}
{2z\bar{z}\,q^-}
}
\int d^2\zperp\,
e^{-i(\koneperp+\ktwoperp)
\cdot\zperp}\,
(\koneperp-\ktwoperp)_i\,
\overline{F}_{-i}
\left(z^-,\zperp\right) \right\}\,,
\end{align}
and the subsequent $y^-$ integration gives
\begin{align}
i \mathcal{M}_{q,L}^{\rm TMD} &=
  \, 4\pi  e \,\epsilon_-(q)   \bar{u}(k_1) \gamma^- v(k_2)\,
  z \bar z \, { {q^- \, \delta(k_g^-) }}
\frac{1}{k^2_{2\perp} + \varepsilon_f^2}
  \frac{1}{k_{1\perp}^2 - k_{2\perp}^2}\int\,dz^-\,
  e^{i z^- \left( \frac{k_{2\perp}^2}{2 z \bar z q^-}  -q^+\right)  }
  \bigg( e^{i  z^- \frac{k_{1\perp}^2 - k_{2\perp}^2}{2k_1^-} } -1 \bigg) 
  \na &
\times \int d^2\zperp\,
e^{-i(\koneperp+\ktwoperp)
\cdot\zperp}\,
(\koneperp-\ktwoperp)_i\,
\overline{F}_{-i}
\left(z^-,\zperp\right) 
\,.
\end{align}
{ At this stage the back-to-back expansion is manifest: the difference
$k_{1\perp}^2-k_{2\perp}^2$ is parametrically small compared with the hard scale $P_\perp^2+\epsilon_f^2$.} Expanding the finite-difference factor to leading nonvanishing order therefore gives
\begin{align}
i \mathcal{M}_{q,L}^{\rm TMD}=
  &\, 8\pi  e \,\epsilon_-(q)   \bar{u}(k_1) \gamma^- v(k_2)\,
  \, {  { \delta(k_g^-) }}
\frac{z \bar z {  {q^-  }}}{(P^2_{\perp} + \varepsilon_f^2)^2}
 \int\,dz^-\,  e^{i z^- k_{P_\perp}^+   }
  \left( i \bar z \;  z^- k_{P_\perp}^+   \right) 
  \na &
\times\int d^2\zperp\,
e^{-i \bDeltaPerp
\cdot\zperp}\,
P_i\,
\overline{F}_{-i}
\left(z^-,\zperp\right) 
\,.
\end{align}

The antiquark contribution similarly gives
\begin{align}
i \mathcal{M}_{\bar{q},L}^{\rm TMD}=
  &\, 8\pi  e \,\epsilon_-(q)   \bar{u}(k_1) \gamma^- v(k_2)\,
  \, {  { \delta(k_g^-) }}
\frac{z \bar z {  {q^-  }}}{(P^2_{\perp} + \varepsilon_f^2)^2}
 \int\,dz^-\,  e^{i z^- k_{P_\perp}^+   }
  \left( i z\; z^- k_{P_\perp}^+   \right) 
  \na &
\times\int d^2\zperp\,
e^{-i \bDeltaPerp
\cdot\zperp}\,
P_i\,
\overline{F}_{-i}
\left(z^-,\zperp\right) 
\,.
\end{align}

The cancellation now becomes transparent.  The quark and antiquark insertions carry the weights $\bar z$ and $z$, respectively, and hence their sum is proportional to
$z+\bar z=1$.  Their combined contribution exactly removes the term generated by differentiating the non-zero-$x$ phase in $\mathcal M_{0,L}$.  The full longitudinal amplitude therefore reduces to
\begin{align}
  i \mathcal{M}_L^{\rm TMD}
 &\approx - 8\pi  e \,\epsilon_-(q) \bar{u}(k_1)\,\gamma^-\,v(k_2)\,  {  { \delta(k_g^-) }}
\frac{z \bar z {  {q^-  }}}{(P^2_{\perp} + \varepsilon_f^2)^2}
      \na & \times \int d\xi^-\, e^{i k_{P_\perp}^+ \xi^-}
 \int d^2 \bperp \,e^{ - i \bDeltaPerp\cdot\bperp }\,
      [\infty,\xi^-]_{\bperp}
      P^i F_{i-}(\xi^-,\bperp)
      [\xi^-,\infty]_{\bperp}
 \,.
\end{align}
The resulting expression is precisely the leading-twist gluon-TMD amplitude for longitudinal dijet production.  Importantly, this result is recovered only after combining the three contributions to the resummed amplitude: the additional term generated by the non-zero-$x$ phase in $\mathcal M_{0,L}$ is exactly canceled by the quark- and antiquark-line contributions.  The resummed amplitude therefore reduces to the expected TMD result in the back-to-back limit.

\subsection{Transverse photon polarization}

The derivation for transverse photon polarization follows the same steps as in the longitudinal case.  We therefore use the reduction of the fermionic endpoint factors derived above and focus on the additional Dirac structure associated with the transverse photon vertex.  For the outgoing quark, this gives
\begin{align}
\label{eq:sym_prop_T}
&\bar{u}(k) \lim_{k^2 \rightarrow 0}\,k^2\,(k | \frac{1}{\Pop^2 + \frac{\sigma F}{2} +i\epsilon }  | y) \gamma^j
\na
&=
\bar{u}(k)  \,e^{ik^- y^+}(\kperp \,|  \,\Bigg[\mathcal{P}\exp\Big(-i \int_{y^-}^{\infty}\,dz^-\,H_{s\sigma}(z^-;\infty;k^-) \Big)[\infty,y^-]\Bigg]R^{-1}(y^-,k^-) \frac{\gamma^-   \slashed{\Pop} }{2k^-}|\, \yperp) \gamma^j.
\end{align}
Applying the analogous reduction to the antiquark propagator and combining the two fermion lines, we obtain the transverse-photon amplitude.  In contrast to the longitudinal case, the fermionic rotations leave a covariant transverse-momentum insertion on each side of the photon vertex:
\begin{align}\label{eq:Amp_T}
  i \mathcal{M}_T& = 
  \, i e \,\epsilon_j(q)
  \int d^4y \, e^{-iq\cdot y}\,e^{i (k_1^- + k_2^-)y^+}\na
  &\times \bar{u}(k_1)\,(\koneperp \,|
  \Bigg[\mathcal{P}\exp\Big(-i \int_{y^-}^{\infty}\,dz^-\,
  H_{s\sigma}(z^-;\infty;k_1^-) \Big)[\infty,y^-]\Bigg]
  R^{-1}(y^-,k_1^-)
  \frac{\gamma^- \slashed \Pop (y^-, k_1^-)}{2k_1^-}
  |\, \yperp)\na &
  \times  \gamma^j
  (\yperp \,|
  \frac{\slashed \Pop(y^-, k_2^-) \gamma^-}{2k_2^-}
  R(y^-, -k_2^-)\Bigg[[y^-,\infty]\mathcal{P}
  \exp\Big(i \int_{y^-}^{\infty}\,dz^-\,
  H_{s\sigma}(z^-;\infty;-k_2^-) \Big)\Bigg]
  |\, -\ktwoperp)\,
 v(k_2)\,.
\end{align}
Here the $k^-$ argument of $\slashed{\Pop}(y^-,k^-)$ specifies the longitudinal momentum entering the corresponding covariant momentum operator. Since $\qperp=0$, the integration over $\yperp$ combines the two covariant transverse momenta at the photon vertex.  Their Dirac structure can be reduced according to
\begin{align}
\label{eq:T-gamma-identity}
\frac{1}{4k_1^-k_2^-}
\gamma^-\slashed{\Pop}(k_1^-)
\gamma^j
\slashed{\Pop}(k_2^-)\gamma^-
&=
\gamma^-
\left[
\left(
\frac{1}{2k_1^-}
+\frac{1}{2k_2^-}
\right)\delta^j_l
+i\left(
\frac{1}{2k_2^-}
-\frac{1}{2k_1^-}
\right)\sigma^{jl}
\right]\Pop^l
=
\frac{1}{2q^-z\bar z},
\gamma^-\mathcal T^{jl}(z)\Pop^l,
\end{align}
where
\begin{equation}
\mathcal T^{jl}(z)
\equiv
\delta^{jl}+i(z-\bar z)\sigma^{jl}.
\end{equation}
The tensor $\mathcal T^{jl}(z)$ collects the spin structure associated with the transverse photon splitting.
The remaining transverse matrix element is treated in the same approximation as in the longitudinal case.  To the accuracy considered here
\begin{align}
\label{T-P-ME}
    (\zoneperp| R^{-1}(y^-,k_1^-) \Pop^l(y^-)  R(y^-, -k_2^-) |\ztwoperp) &\approx 
    [\zoneperp, \ztwoperp]_{y^-}
    (\zoneperp| \mathrm{p}^l e^{i \frac{y^- p^2}{2\kappa^-}} | \ztwoperp)
    \na &=  [\zoneperp, \ztwoperp]_{y^-}
    \int { \bar d}^2 \pperp \; p^l
     e^{i (\zoneperp - \ztwoperp) \cdot\pperp  }  e^{i \frac{y^-\,p_{\perp}^2}{2 \kappa^-}  }\,.
\end{align}

\subsubsection{Semi-infinite dipole contribution}
We first consider the contribution obtained by setting the ordered evolution operators to unity.  The structure is closely related to its longitudinal counterpart, with the transverse polarization introducing an additional transverse momentum $p^l$ together with the spin tensor $\mathcal T^{jl}(z)$.  Using Eqs.~\eqref{eq:T-gamma-identity} and \eqref{T-P-ME}, we obtain
\begin{align}
\label{eq:AmplitudeM2-transverse}
 i\mathcal M_{0,T}
 &= \frac{i e\,\epsilon_j(q)}{2q^-z\bar z}\,
 \bar u(k_1)\gamma^-\mathcal T^{jl}(z)v(k_2)
 \int d^2\zoneperp d^2\ztwoperp\,
 e^{-i\koneperp\cdot\zoneperp-i\ktwoperp\cdot\ztwoperp}
 \int {\bar d}^{\,2}\pperp\;p^l
 e^{i(\zoneperp-\ztwoperp)\cdot\pperp}
 \na & \times
 \int dy^-dy^+\,e^{ik^+_{\pperp}y^-}e^{-ik_g^-y^+}
 S(y^-;\zoneperp,\ztwoperp),
\end{align}
where
\begin{align}
 S(y^-;\zoneperp,\ztwoperp)
 &\equiv[\infty,y^-]_{\zoneperp}
 [\zoneperp,\ztwoperp]_{y^-}
 [y^-,\infty]_{\ztwoperp},
\end{align}
and $k^+_{\pperp}$ is defined in Eq.~\eqref{Eq:k^+}.  Apart from the transverse-photon numerator, Eq.~\eqref{eq:AmplitudeM2-transverse} contains the same resummed transverse propagation and staple structure as the longitudinal amplitude.  

\subsubsection{Contributions from the evolution operators}

We next restore the contributions from the ordered evolution operators.  Their scalar part is independent of the photon polarization, so the corresponding field-strength insertions have the same operator structure as in the longitudinal case.  The transverse polarization modifies only the numerator through the factor $p^l\mathcal T^{jl}(z)/(2q^-z\bar z)$.  The quark-line contribution therefore takes the form
\begin{align}
\label{eq:AmplitudeMq-T}
 i\mathcal M_{q,T}
 &=\frac{i e\,\epsilon_j(q)}{2q^-z\bar z}
 \bar u(k_1)\gamma^-\mathcal T^{jl}(z)v(k_2)
 \int dy^-dy^+\,e^{-iq^+y^-}e^{-ik_g^-y^+}
 \frac{1}{k_{1\perp}^2-k_{2\perp}^2}\int d^2\zoneperp d^2\ztwoperp\,
 e^{-i\koneperp\cdot\zoneperp-i\ktwoperp\cdot\ztwoperp}
 \na & \times
 \int_{y^-}^{\infty}d\xi^-
 \bigg(e^{i\xi^-\frac{k_{1\perp}^2-k_{2\perp}^2}{2k_1^-}}-1\bigg)
 \int {\bar d}^{\,2}\pperp\;p^l
 e^{i(\zoneperp-\ztwoperp)\cdot\pperp}
 e^{i\frac{y^-p_\perp^2}{2\kappa^-}}
 (k_1-k_2)_i\,
 \bar F_{-i}(\xi^-,\zoneperp)
 S(y^-;\zoneperp,\ztwoperp).
\end{align}

The antiquark-line contribution follows from Eq.~\eqref{eq:AmplitudeMq-T} under the interchange
\begin{equation}
\label{eq:AmplitudeMqbar-T}
k_1,z,\zoneperp
 \leftrightarrow 
k_2,\bar z,\ztwoperp,
\end{equation}
with the rotated field-strength insertion placed on the right of the staple.  As in the longitudinal case, the term proportional to $\sigma^{ij}F_{ij}$ is suppressed in both the high-energy and TMD power countings and is therefore omitted.  Together, the quark- and antiquark-line contributions account for the evolution-operator corrections to the transverse amplitude at the accuracy considered here.

\subsubsection{Eikonal limit}

We first verify that the transverse amplitude reproduces the expected high-energy result.  Since the dependence on $y^-$ in Eq.~\eqref{eq:AmplitudeM2-transverse} is carried by the same staple integral as for longitudinal polarization, its eikonal reduction can be taken over directly from the preceding subsection.  At leading order,
\begin{align}
 \int dy^-e^{ik^+_{p_\perp}y^-}S(y^-)
 &=\frac{i}{k^+_{p_\perp}}
 \left(1-S(-\infty)\right)+\mathcal O(\text{sub-eikonal}),
\end{align}
Substituting this result into Eq.~\eqref{eq:AmplitudeM2-transverse} gives
\begin{align}
\label{eq:AmplitudeM0-transverse-eik}
 i\mathcal M_{T}^{\mathrm{eik}}
 &=-e\,\epsilon_j(q)\,
 \bar u(k_1)\gamma^-\mathcal T^{jl}(z)v(k_2)\,
 \bar\delta(k_g^-)
 \int d^2\zoneperp d^2\ztwoperp\,
 e^{-i\koneperp\cdot\zoneperp-i\ktwoperp\cdot\ztwoperp}
 \na & \times
 \int {\bar d}^{\,2}\pperp\,
 \frac{p^l e^{i(\zoneperp-\ztwoperp)\cdot\pperp}}
 {p_\perp^2+\epsilon_f^2}
 \left(1-S(-\infty;\zoneperp,\ztwoperp)\right).
\end{align}
The remaining Fourier transform differs from the longitudinal case by the factor of $p^l$ in the numerator.  It can be evaluated as
\begin{align}
 \int {\bar d}^{\,2}\pperp\,
 \frac{p^l e^{i\rperp\cdot\pperp}}{p_\perp^2+\epsilon_f^2}
 =\frac{i\epsilon_f}{2\pi}\frac{r^l}{r_\perp}
 K_1(\epsilon_f r_\perp).
\end{align}
The Fourier transform therefore gives the expected $K_1$ kernel for transverse photon polarization.  Equation~\eqref{eq:AmplitudeM0-transverse-eik} then reproduces the transverse-photon eikonal amplitude.

The resulting amplitude agrees with the transverse projection of Ref.~\cite{Kar:2026vzk}.  When the transverse background field is set to zero, the transverse segment of the staple becomes trivial and the result reduces to the conventional CGC dipole expression of Refs.~\cite{Dominguez:2011wm,Dominguez:2011br}. As in the longitudinal case, we see that the contributions from the operators Fig.~\ref{fig:q_con} and Fig.~\ref{fig:aq_con}   do not contribute at the eikonal order.

\subsubsection{TMD limit}
We finally consider the back-to-back limit. Using the transverse coordinates introduced in the longitudinal analysis, we expand the staple to first order in the relative separation $\rperp$:
\begin{align}
 S(y^-;\zoneperp,\ztwoperp)
 &=1+i r_\perp^i\int_{y^-}^{\infty}d\xi^-\,
 [\infty,\xi^-]_{\bperp}F_{i-}(\xi^-,\bperp)
 [\xi^-,\infty]_{\bperp}+\mathcal O(r_\perp^2).
\end{align}
The unit term is proportional to $\delta(k^+_{\pperp})$ and does not contribute for $k^+_{\pperp}>0$.  Keeping the term linear in $\rperp$ and performing the $y^-$ and $\rperp$ integrations, the semi-infinite dipole contribution becomes
\begin{align}
\label{eq:AmpM0-T-TMD-raw}
 i\mathcal M_{0,T}^{\mathrm{TMD}}
 &=\, { \bar \delta(k_g^-)}\,
 \frac{e\,\epsilon_j(q)}{2q^-z\bar z}\,
 \bar u(k_1)\,\gamma^-\,\mathcal T^{jl}(z)v(k_2)\,
 \int d\xi^-\int d^2\bperp\,e^{-i\bDeltaPerp\cdot\bperp}
 \na & \times
 \frac{\partial}{\partial P_i}
 \left(\frac{P^l e^{ik^+_{P_\perp}\xi^-}}
 {k^+_{P_\perp}}\right)
 [\infty,\xi^-]_{\bperp}F_{i-}(\xi^-,\bperp)
 [\xi^-,\infty]_{\bperp}\,.
\end{align}
In contrast to the longitudinal case, the transverse-photon numerator introduces an additional factor of $P^l$.  The derivative generated by the small-$\rperp$ expansion therefore acts both on the prefactor $P^l/k^+_{\Pperp}$ and on the non-zero-$x$ phase:
\begin{align}
 \frac{\partial}{\partial P_i}
 \left(\frac{P^l e^{ik^+_{P_\perp}\xi^-}}{k^+_{P_\perp}}\right)
 &=e^{ik^+_{P_\perp}\xi^-}
 \left[
 \frac{\partial}{\partial P_i}\left(\frac{P^l}{k^+_{P_\perp}}\right)
 +i\xi^-\frac{P^l}{k^+_{P_\perp}}
 \frac{\partial k^+_{P_\perp}}{\partial P_i}
 \right].
\end{align}
The first term gives the transverse hard tensor of the leading-twist TMD amplitude.  The second originates solely from the $P_\perp$ dependence of the non-zero-$x$ phase and, as in the longitudinal calculation, must be combined with the contributions {from Fig.~\ref{fig:q_con} and Fig.~\ref{fig:aq_con}}.

Expanding the quark- and antiquark-line terms in the same limit gives the relative weights $\bar z$ and $z$, respectively. Their sum is
\begin{align}
\label{eq:AmpMqa-T-TMD}
 i\mathcal M_{q,T}^{\mathrm{TMD}}+i\mathcal M_{\bar q,T}^{\mathrm{TMD}}
 &=- { \bar\delta(k_g^-)}
 \frac{e\,\epsilon_j(q)}{2q^-z\bar z}
 \bar u(k_1)\gamma^-\mathcal T^{jl}(z)v(k_2)
 \int d\xi^-\int d^2\bperp\,e^{-i\bDeltaPerp\cdot\bperp}
 \na & \times e^{ik^+_{P_\perp}\xi^-}
 i\xi^-\frac{P^l}{k^+_{P_\perp}}
 \frac{\partial k^+_{P_\perp}}{\partial P_i}
 [\infty,\xi^-]_{\bperp}F_{i-}(\xi^-,\bperp)
 [\xi^-,\infty]_{\bperp}.
\end{align}
This contribution exactly cancels the phase-derivative term in Eq.~\eqref{eq:AmpM0-T-TMD-raw}.  The complete leading-twist transverse amplitude is therefore
\begin{align}
\label{eq:AmpMqa-T-TMD-final}
 i\mathcal M_T^{\mathrm{TMD}}
 &= { \bar \delta(k_g^-)}\,e\,\epsilon_j(q)\,
 \bar u(k_1)\gamma^-\mathcal T^{jl}(z)v(k_2)
 \int d\xi^-e^{ik^+_{P_\perp}\xi^-}
 \int d^2\bperp\,e^{-i\bDeltaPerp\cdot\bperp}
 \na & \times
 \left[
 \frac{\delta^{il}}{P_\perp^2+\epsilon_f^2}
 -\frac{2P^iP^l}{(P_\perp^2+\epsilon_f^2)^2}
 \right]
 [\infty,\xi^-]_{\bperp}F_{i-}(\xi^-,\bperp)
 [\xi^-,\infty]_{\bperp}.
\end{align}
This result has the expected leading-twist gluon-TMD operator structure, with the transverse-photon dependence contained in $\mathcal T^{jl}(z)$ and the hard transverse tensor.  As for longitudinal polarization, the cancellation of the phase-derivative term requires the semi-infinite dipole and {explicit field-strength} contributions to be combined before taking the TMD limit.  The longitudinal and transverse results thus provide complementary checks that the resummed non-zero-$x$ amplitude reproduces both the eikonal and leading-twist TMD limits.


\section{Connection to iTMD}
\label{sec:itmd_rep}

\begin{figure}[t]
    \centering
    \includegraphics[width=0.5\linewidth]{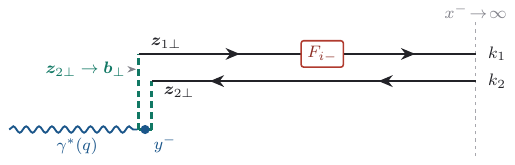}
    \caption{
    \justifying
    Small-$x$ iTMD dijet amplitude.  The transverse separation of the pair is not expanded, so that the full transverse kinematics is retained, and the field strength is instead placed at the common transverse position  $\bperp$ by setting $\ztwoperp\to\bperp$. }
     \label{fig:itmd_con}
\end{figure}

The preceding analysis shows that the resummed dijet amplitude reproduces both the eikonal small-$x$ result and the leading-twist TMD limit.  This raises a natural question: whether the full non-zero-$x$ expression can be reorganized into an iTMD-like form, in which the transverse kinematics are resummed into the hard factor while the target dependence is encoded in a TMD operator.  We examine this possibility using the semi-infinite dipole longitudinal contribution $\mathcal M_{0,L}$.  This term is sufficient to expose the relevant structure.

The standard iTMD construction isolates the field-strength insertion at a common transverse position while retaining the full dependence on the hard transverse momenta.  
We therefore use Eq.~\eqref{eq:final-result2dtransporter} to rewrite the transverse Wilson-line structure as
\begin{align}
&[\infty,y^-]_{\zoneperp}[\zoneperp,\ztwoperp]_{y^-}\,\na &=\,[\infty,y^-]_{\ztwoperp}\,-i(z_{2\perp}^i - z_{1\perp}^{i})\,\int\,\frac{d^2 \kperp}{(2\pi)^2}\,\int\,d^2 \bperp\,e^{-i \kperp \cdot (\bperp - \zoneperp)}\,\frac{e^{i\kperp \cdot (\ztwoperp - \zoneperp)} - 1}{\kperp \cdot (\ztwoperp - \zoneperp)}\,\partial_i([\infty, y^-]_{\bperp} [\bperp,\ztwoperp]_{y^-}).
\end{align}
Substituting this relation into the semi-infinite dipole amplitude gives
\begin{align}\label{eq:AmplitudeM4}
  i \mathcal{M}_{0,L}^{\rm iTMD}
 & = i e \,\epsilon_\rho(q) \int d^2 \zoneperp\, d^2 \ztwoperp  e^{-i \koneperp\cdot\zoneperp -i \ktwoperp\cdot\ztwoperp} 
  \int { \bar d}^2 \pperp  
     e^{i (\zoneperp - \ztwoperp) \cdot\pperp  }
  \int dy^- dy^+   e^{i k_{p_\perp}^+   y^-}  e^{- i k_g^- y^+} \bar{u}(k_1)\,\gamma^\rho\,v(k_2)
  \na & \times 
  \bigg(1-i(z_{2\perp}^i \!-\! z_{1\perp}^{i})\int\frac{d^2 \kperp}{(2\pi)^2} \int d^2 \bperp e^{-i \kperp \cdot (\bperp - \zoneperp)} \frac{e^{i\kperp \cdot (\ztwoperp - \zoneperp)} \!-\! 1}{\kperp \cdot (\ztwoperp - \zoneperp)} \partial_i([\infty, y^-]_{\bperp}  [\bperp,\ztwoperp]_{y^-} )[y^-,\infty]_{\ztwoperp}  \bigg)\
 \,.
\end{align}

The transverse derivative generates a field-strength insertion along the longitudinal 
\begin{align}
    \partial_i([\infty, y^-]_{\bperp} [\bperp,\ztwoperp]_{y^-}) 
    = 
   -i  \int_{y^-}^\infty dz^- \bar F_{-i}(z^-, \bperp) [\infty, y^-]_{\bperp}.
\end{align}
Multiplying by the remaining Wilson line therefore gives
\begin{align}
    \partial_i([\infty, y^-]_{\bperp} [\bperp,\ztwoperp]_{y^-})[y^-,\infty]_{\ztwoperp}   = -i \int_{y^-}^\infty dz^- \bar F_{-i} (z^-, \bperp) [\infty, y^-]_{\bperp}  [\bperp,\ztwoperp]_{y^-} [y^-,\infty]_{\ztwoperp}.
\end{align}

The same procedure can be applied to the remaining Wilson-line product  $[\bperp,\ztwoperp]_{y^-}[y^-,\infty]_{\ztwoperp}$.  Here we keep only the contribution in which the field strength is evaluated at the common transverse position $\bperp$, as relevant for the TMD-like structure considered below.  Terms that retain fields at different transverse positions belong to the higher-twist sector and are not considered here.  We therefore set $\ztwoperp\to\bperp$ in this contribution and obtain
\begin{align}
    \partial_i([\infty, y^-]_{\bperp} [\bperp,\ztwoperp]_{y^-})[y^-,\infty]_{\ztwoperp}   \approx  -i \int_{y^-}^\infty dz^- \bar F_{-i} (z^-, \bperp).
\end{align}
The background-independent term in Eq.~\eqref{eq:AmplitudeM4} has no support in the physical scattering amplitude and will be omitted.  The projected contribution then becomes
\begin{align}\label{eq:AmplitudeM5}
  i \mathcal{M}_{0,L}^{\rm iTMD}
 & = - i e \,\epsilon_\rho(q) \int \,d^2 \bperp \int { \bar d}^2 \pperp  \, \bar d^2 \kperp  \;
 \int dy^- dy^+ \, \int_{y^-}^\infty dz^-  
  \, e^{i k_{p_\perp}^+   y^-}\,e^{- i k_g^- y^+} \bar F_{-i} (z^-, \bperp) \bar{u}(k_1)\,\gamma^\rho\,v(k_2)
  \na &
   \int d^2 \zoneperp \,d^2 \ztwoperp e^{-i \koneperp\cdot\zoneperp -i \ktwoperp\cdot\ztwoperp}\,
    \,
     e^{i (\zoneperp - \ztwoperp) \cdot\pperp  }
  (z_{2\perp}^i - z_{1\perp}^{i}) \,e^{-i \kperp \cdot (\bperp - \zoneperp)}\,\frac{e^{i\kperp \cdot (\ztwoperp - \zoneperp)} - 1}{\kperp \cdot (\ztwoperp - \zoneperp)}
 \,.
\end{align}

The remaining transverse integrations can be carried out by using
\begin{align}
    \frac{e^{i\kperp \cdot (\ztwoperp - \zoneperp)} - 1}{\kperp \cdot (\ztwoperp - \zoneperp)}
     = i \int_0^1 dt\, e^{i t \kperp \cdot  (\ztwoperp - \zoneperp)},
\end{align}
The factor $(z_{2\perp}^{i}-z_{1\perp}^{i})$ can then be converted into a derivative with respect to $\pperp$.  Performing the $\zoneperp$ and $\ztwoperp$ integrations gives
\begin{align}
&\int d^2\zoneperp d^2\ztwoperp\,
e^{-i\koneperp\cdot\zoneperp-i\ktwoperp\cdot\ztwoperp}
(z_{2\perp}^{i}-z_{1\perp}^{i})
e^{i(\zoneperp-\ztwoperp)\cdot\pperp}
e^{-i\kperp\cdot(\bperp-\zoneperp)}
\frac{
e^{i\kperp\cdot(\ztwoperp-\zoneperp)}-1
}{
\kperp\cdot(\ztwoperp-\zoneperp)
}
\na
&\qquad =
-\frac{\partial}{\partial p^i}
\left[
\int_0^1dt\,
\bar\delta^{(2)}
(\koneperp+\ktwoperp-\kperp)
\,
\bar\delta^{(2)}
\left(
\bar t\,\ktwoperp-t\,\koneperp-\pperp
\right)
\right]
e^{-i\kperp\cdot\bperp},
\label{eq:iTMD-transverse-integrals}
\end{align}
The longitudinal integrations can likewise be performed by interchanging their order.  Using
\begin{align}
    \int dy^- \int_{y^-}^{\infty} d z^- e^{-i k_{p_\perp}^+ y^-} G(z^-) 
    =
    \frac{i}{k_{p_\perp}^+} \int dz^-  e^{-i k_{p_\perp}^+ z^-} G(z^-).
\end{align}
After performing the transverse-coordinate and longitudinal integrations, the amplitude becomes
\begin{align}\label{eq:AmplitudeM5_after_integrals}
  i \mathcal{M}_{0,L}^{\rm iTMD}
 & = - e \,\epsilon_\rho(q) \int \,d^2 \bperp \int { \bar d}^2 \pperp  \, \bar d^2 \kperp  \;
  \int dy^+ \,dz^-  
  \, 
  \frac{e^{i k_{p_\perp}^+   z^-}}{k_{p_\perp}^+}\,e^{- i k_g^- y^+} \bar F_{-i} (z^-, \bperp) \bar{u}(k_1)\,\gamma^\rho\,v(k_2)
  \na &
  \frac{\partial}{\partial p^i} \left(   \int_0^1 dt\,  \bar \delta^{(2)}(\koneperp + \ktwoperp - \kperp)  \bar \delta^{(2)}\left( \bar t \ktwoperp - t \koneperp   - \pperp  \right)  \right)
   e^{-i \kperp \cdot \bperp}
 \,.
\end{align}
We now integrate by parts with respect to $\pperp$, transferring the transverse derivative from the delta function to the factor $e^{ik^+_{p_\perp}z^-}/k^+_{p_\perp}$.  The remaining $\pperp$ and $\kperp$ integrations are then fixed by the transverse delta functions.  For longitudinal polarization, this gives
\begin{align}\label{eq:AmplitudeM0_final}
  i \mathcal{M}_{0,L}^{\rm iTMD}
 & =  e \,\epsilon_-(q)
  \bar{u}(k_1)\,\gamma^-\,v(k_2) { { \delta(k_g^-) }}
 \int \,d^2 \bperp \int_0^1 dt\,  \;
  \int  \,dz^-  
  \, 
  \frac{\partial}{\partial p^i}  \left( \frac{e^{i k_{p_\perp}^+   z^-}}{k_{p_\perp}^+} \right) \bar F_{-i} (z^-, \bperp) 
   e^{-i \kqperp \cdot \bperp}\,,
\end{align}
where
\begin{align}
    \pperp & = - t\, \koneperp + \bar t\, \ktwoperp\,.
\end{align}

In the eikonal limit, the longitudinal phases collapse to their small-$x$
form and $i\mathcal M_{0,L}^{\rm iTMD}$ reproduces the corresponding iTMD amplitude \cite{Altinoluk:2019wyu}.  At
non-zero $x$, however, $\pperp=-t\koneperp+\bar t\ktwoperp$ varies continuously
with $t\in[0,1]$.  The frequency $k^+_{p_\perp}$ in
Eq.~\eqref{eq:AmplitudeM0_final} therefore also varies with $t$.  The amplitude
contains a continuum of phases $\exp(i k^+_{p_\perp}z^-)$ rather than the
single longitudinal phase associated with a conventional gluon TMD at fixed
$x$.  Consequently, this contribution cannot be written as an iTMD hard factor
times a standard non-zero-$x$ TMD operator.  This obstruction does not affect
either limiting check established above. This suggests that our operator basis  
is more general than the standard iTMD operator structure.

\section{Conclusions}
\label{sec:summary}

The aim of this work was to investigate whether the all-twist eikonal regime and the back-to-back arbitrary-$x$ leading-twist TMD regime of the DIS dijet production can be bridged through an underlying unifying operator basis. To this end, we reorganized the quark propagator in the background gluon field through an endpoint transformation. The resulting propagator resums transverse kinematic corrections in the endpoint rotations while its ordered Hamiltonian is organized by explicit field strengths. This form makes the two expansions transparent and implements a simple accuracy criterion: we retain every term required by either the leading-eikonal result at all twists or the leading-twist result at non-zero $x$, and neglect terms that are simultaneously sub-eikonal and higher twist suppressed. This enabled us to express the expansion of the background-field propagator in an operator basis containing a semi-infinite dipole operator and corrections containing field-strength insertions into the Wilson lines. As we noted in Appendix \ref{app:reduction_F} the semi-infinite dipole operator in Fig.~\ref{fig:sym_con} can be expressed in terms of operators Fig.~\ref{fig:q_con} and Fig.~\ref{fig:aq_con}. 

We then derived the DIS dijet amplitudes for longitudinally and transversely polarized photons in this operator basis. The resulting amplitudes pass two independent checks. In the $x\to0$ eikonal limit, the contribution of the semi-infinite dipole operator dominates and reduces to the standard eikonal CGC expressions for the longitudinal and the transverse photon polarizations, including the $K_0$ and $K_1$ photon wave-function kernels. In the back-to-back limit, the amplitudes coincide with the leading-twist gluon-TMD result at arbitrary $x$ for both photon polarizations. The importance of the operators containing field-strength insertions becomes clear while reproducing this limit. Accounting for these operators becomes a necessity to cancel the contributions of the derivatives of the longitudinal phase at $x>0$ associated with the semi-infinite dipole operator. Together these results show that our operator basis keeps the terms needed in either limit, while omitting corrections that are simultaneously sub-eikonal and beyond leading twist.

As a first proof-of-principle study, this work restricts the new operator basis to the operators necessary to demonstrate explicit matching to the leading-eikonal and leading-twist accuracies in the respective limits. Moreover, since our background-field propagator is general, there is no obvious technical obstruction to carrying out its systematic expansion in terms of a broader operator basis that also accurately captures sub-eikonal and twist-suppressed corrections.

Further, we found that the DIS dijet amplitude in terms of our new operator basis admits a natural connection to the eikonal iTMD factorization. We demonstrated that our transverse resummation generates a continuous family of longitudinal phases. In the eikonal limit these phases become indistinguishable and the standard iTMD structure is recovered. However, it must be emphasized that at non-zero $x$ these longitudinal phases cannot be absorbed into a single conventional TMD operator. This suggests that extension of iTMD-type factorization to capture sub-eikonal corrections might be difficult.

{The present study is restricted to unpolarized DIS. An important open question is whether this operator basis suffices also for polarized DIS, where the leading CGC contribution is sub-eikonal. We leave this investigation for a future study.}

\acknowledgements
We thank Andrey Tarasov and Yacine Mehtar-Tani for illuminating discussions. This work is supported by the U.S. Department of Energy, Office of Science, Office of Nuclear Physics through Contract Nos.~DE-SC0012704 and DE-SC0020081, the Saturated Glue (SURGE) Topical Collaboration in Nuclear Theory, and the United States-Israel Binational Science Foundation grant \#2022132. 
\appendix

\section{The Duhamel formula} 
\label{sec:duhamel}

Let \(U(a,b)\) denote the evolution operator from \(b\) to \(a\), satisfying
\begin{equation}
  i\frac{\partial}{\partial a}U(a,b)
  =
  H(a)U(a,b),
  \qquad
  U(b,b)=1.
\end{equation}
The formal solution is
\begin{equation}
  U(a,b)
  =
  \mathcal P
  \exp\left[
    -i\int_b^a d\xi\,H(\xi)
  \right],
\end{equation}
where \(\mathcal P\) orders later values of \(\xi\) to the left.

We consider Hamiltonians that can be separated into an exactly solvable part
and a perturbation,
\begin{equation}
  H(a)=H_0+V(a),
\end{equation}
where for simplicity \(H_0\) is independent of \(a\). The free evolution is
\begin{equation}
  U_0(a,b)=e^{-i(a-b)H_0}.
\end{equation}

The full evolution can then be written as free propagation with interactions
generated by \(V\).  Duhamel's formula gives
\begin{equation}
  U(a,b)
  =
  U_0(a,b)
  -
  i\int_b^a d\xi\,
  U_0(a,\xi)V(\xi)U(\xi,b).
\end{equation}

Duhamel's formula is exact.  Its first-order expansion follows by replacing
\(U(\xi,b)\) in the integrand with the free evolution \(U_0(\xi,b)\):
\begin{equation}
  U(a,b)
  =
  U_0(a,b)
  -
  i\int_b^a d\xi\,
  U_0(a,\xi)V(\xi)U_0(\xi,b)
  +
  O(V^2).
\end{equation}
Equivalently,
\begin{equation}
\label{eq:duhamel_o1}
  U(a,b)
  =
  e^{-i(a-b)H_0}
  -
  i\int_b^a d\xi\,
  e^{-i(a-\xi)H_0}
  V(\xi)
  e^{-i(\xi-b)H_0}
  +
  O(V^2).
\end{equation}
The interpretation is immediate: the system evolves freely from
\(b\) to \(\xi\), interacts through \(V(\xi)\), and then evolves freely
from \(\xi\) to \(a\). The integration sums over all possible interaction
points \(\xi\).

Iterating Duhamel's formula generates the full ordered series.  The first two
nontrivial orders are
\begin{align}
\label{eq:duhamel_o2}
  U(a,b)
  &=
  U_0(a,b)
  -
  i\int_b^a d\xi_1\,
  U_0(a,\xi_1)V(\xi_1)U_0(\xi_1,b)
  \nonumber\\
  &\quad
  +
  (-i)^2
  \int_b^a d\xi_1
  \int_b^{\xi_1} d\xi_2\,
  U_0(a,\xi_1)V(\xi_1)
  U_0(\xi_1,\xi_2)V(\xi_2)
  U_0(\xi_2,b)
  +\cdots .
\end{align}

\section{Useful relations}

The transverse derivative of a longitudinal Wilson line is
\begin{align}
  \partial_k [x^-, y^-]_{\zperp}  &= i A_k(x^-, \zperp)  [x^-, y^-]_{\zperp}  -  i [x^-, y^-]_{\zperp} A_k(y^-, \zperp) \na &- i \int_{y^-}^{x^-} d\zeta^- [x^-, \zeta^-]_{\zperp} F_{-k} (\zeta^-,\zperp) [\zeta^-, y^-]_{\zperp}\,.
  \label{eq:derWL}
\end{align}
For a straight Wilson line along the minus direction, one also finds
\begin{align}
  \label{eq:wilson_comm}
  [x^-,y^-]\,{\rm P}_i(y^-)
  = {\rm P}_i(x^-)\,[x^-,y^-]
  - \int_{y^-}^{x^-} dz^-\,
    [x^-,z^-]\,F_{-i}(z^-)\,[z^-,y^-] \,.
\end{align}

In the amplitude, structures involving $D_iF_{-i}$ can be reduced to terms
containing one or two field strengths.  Integration by parts, together with
Eq.~\eqref{eq:derWL}, gives
\begin{align}
 \int dz\, e^{i k \cdot  z } \, 
     [\infty,z^-]_{\zperp} D_j F_{-i}(z^-, \zperp)\,[z^-,\infty]_{\zperp} &=\int dz\, e^{i k \cdot  z } 
     \Bigl\{
       -i  k_j \,[\infty,z^-]_{\zperp} F_{-i}(z^-,\zperp)[z^-,\infty]_{\zperp}
   \na     
& +   i   \int_{z^-}^\infty dz'^- \,[\infty,z'^-]_{\zperp} F_{-j} (z'^-,\zperp) [z'^-,z^-] F_{-i} (z^-,\zperp) [z^-,\infty]_{\zperp}  \notag\\
& -i   \int_{z^-}^{\infty} dz'^- \,[\infty,z^-]_{\zperp}  F_{-i}(z^-,\zperp)  [z^-, z'^-] F_{-j}(z'^-,\zperp) [z'^-,\infty]_{\zperp} 
     \Bigr\}\,.  
\label{eq:DF-simplify}
\end{align}

\section{Matrix Element of the \texorpdfstring{$\{P_i,F_{-i}\}$}{\{P,F\}} Insertion}
\label{app:MatrixElemPF}
We begin by evaluating the anticommutator
$\{{\mathrm P}_i,F_{-i}\}$ before introducing transverse position states:
\begin{align}
  & (\koneperp \,|
  [\infty,z^-] \{{\mathrm P}_i(z^-), F_{-i}(z^-)\} [z^-,y^-] 
  R^{-1}\left(y^-, \kappa^- \right) 
  [y^-,\infty] |\, -\ktwoperp)\na
  &= 
  (k_1 - k_2)_i \; (\koneperp \,|
  \bar F_{-i}(z^-)  S_R(y^-) |\, -\ktwoperp) -
  (\koneperp \,| \int^\infty_{z^-} d\zeta^- \bar F_{-i}(\zeta^-) \bar F_{-i}(z^-)
   S_R(y^-) |\, -\ktwoperp) \na 
  & +
  (\koneperp \,| \int^{z^-}_{y^-} d\zeta^- \bar F_{-i}(z^-) \bar F_{-i}(\zeta^-)
   S_R(y^-) |\, -\ktwoperp) -
  (\koneperp \,|  \bar F_{-i}(z^-) 
  S_R(y^-) \int^{\infty}_{y^-} d\zeta^- \bar F_{- i} (\zeta^-) |\, -\ktwoperp),
\end{align}
where 
\begin{align}
    S_R(y^-) = [\infty, y^-]
  R^{-1}\left(y^-, \kappa^-\right) 
  [y^-,\infty]\,.  
\end{align}

Inserting transverse resolutions of the identity produces the matrix element
\begin{align}
    (\zoneperp| S_R(y^-) |\ztwoperp)  \approx  [\infty, y^-]_{\zoneperp} [\zoneperp, \ztwoperp]_{y^-} [y^-,\infty]_{\ztwoperp} 
    \int { \bar d}^2 \pperp \;
     e^{i (\zoneperp - \ztwoperp) \cdot\pperp  }  e^{i \frac{y^-\, p_{\perp}^2}{ 2 \kappa^-} }\,,
\end{align} 
where Eq.~\eqref{eq:z1_Free_Prop_z2} was used.

Combining these relations gives

\begin{align}
  & (\koneperp \,|
  [\infty,z^-] \{{\mathrm P}_i(z^-), F_{-i}(z^-)\} [z^-,y^-] 
  R^{-1}\left(y^-, \kappa^- \right) 
  [y^-,\infty] |\, -\ktwoperp)\na
  &= 
  \int d^2 \zoneperp d^2 \ztwoperp  
  e^{ - i \koneperp\cdot\zoneperp  - i \ktwoperp\cdot\ztwoperp }
  \int { \bar d}^2 \pperp \;
     e^{i (\zoneperp - \ztwoperp) \cdot\pperp  }  e^{i \frac{y^-\, p_{\perp}^2} {2\kappa^-} } \na
 & ~~~~~\times \Bigg\{
  (k_1 - k_2)_i \; 
  \bar F_{-i}(z^-, \zoneperp ) [\infty, y^-]_{\zoneperp} [\zoneperp, \ztwoperp]_{y^-} [y^-,\infty]_{\ztwoperp}  \na 
  & ~~~~~ -
   \int^\infty_{z^-} d\zeta^- \big\{ \bar F_{-i}(\zeta^-,\zoneperp) , \bar F_{-i}(z^-,\zoneperp) \big\}
   [\infty, y^-]_{\zoneperp} [\zoneperp, \ztwoperp]_{y^-} [y^-,\infty]_{\ztwoperp}  \na 
  & ~~~~~ +
   \bar F_{-i}(z^-,\zoneperp) \int^{\infty}_{y^-} d\zeta^-  \bar F_{-i}(\zeta^-,\zoneperp)
    [\infty, y^-]_{\zoneperp} [\zoneperp, \ztwoperp]_{y^-} [y^-,\infty]_{\ztwoperp} 
   \na 
  &~~~~~  -
  \bar F_{-i}(z^-,\zoneperp) 
  [\infty, y^-]_{\zoneperp} [\zoneperp, \ztwoperp]_{y^-} [y^-,\infty]_{\ztwoperp} \int^{\infty}_{y^-} d\zeta^- \bar F_{- i} (\zeta^-, \ztwoperp) \Bigg\}
  .
\end{align}

The full expression is multiplied by $z^-$ and is therefore already
sub-eikonal.  According to our accuracy criterion, only its leading-twist term
must be retained; the multi-field terms displayed above are additionally
higher twist.  Hence

\begin{align}
  & (\koneperp \,|
  [\infty,z^-] \{{\mathrm P}_i(z^-), F_{-i}(z^-)\} [z^-,y^-] 
  R^{-1}\left(y^-, \kappa^- \right) 
  [y^-,\infty] |\, -\ktwoperp)
  = \na
  & ~~~~
  \int d^2 \zoneperp d^2 \ztwoperp  
  e^{ - i \koneperp\cdot\zoneperp  - i \ktwoperp\cdot\ztwoperp }
  \int { \bar d}^2 \pperp \;
     e^{i (\zoneperp - \ztwoperp) \cdot\pperp  }  e^{i \frac{y^-\, p_{\perp}^2}{2 \kappa^-} } 
     \na & ~~~~ ~~~~ \times 
  (k_1 - k_2)_i \; 
  \bar F_{-i}(z^-, \zoneperp ) [\infty, y^-]_{\zoneperp} [\zoneperp, \ztwoperp]_{y^-} [y^-,\infty]_{\ztwoperp}.
\end{align}
%
%
%
%
\section{Expansion of transverse exponential}
\label{sec:TEexpansion}
We begin with the transverse matrix element
\begin{equation}
\mathcal{K}_s(\xperp,\yperp)\,=\,(\xperp|e^{i s \Ptop^2}|\yperp)\,.
\end{equation}
Our goal is to expand it around the straight link connecting $\xperp$ and
$\yperp$.

It is convenient to introduce an auxiliary object,  $\tilde{\mathcal{K}}_s(\xperp,\yperp) = \,  [\yperp,\xperp]\, \mathcal{K}_s(\xperp,\yperp)$, which can be transformed into 
\begin{equation}
\begin{aligned}
&\tilde{\mathcal{K}}_s(\xperp,\yperp)\,=\,[\yperp,\xperp]\,(\xperp|e^{i s \Ptop^2}|\yperp)\,[\yperp,\yperp]\,
\,=\,(\xperp|\hat U_y^{\dagger}\,e^{i s \Ptop^2}\,\hat U_y|\yperp) 
= (\xperp| e^{is \hat U_y^{\dagger} \Ptop^2 \hat U_y}|\yperp),
\end{aligned}
\label{eq:Ktilde}
\end{equation}
where $\hat U_y=[\hatxperp,\yperp]$ and $\hatxperp$ is the transverse position
operator.  Hats distinguish operators from their matrix elements.

To determine the transformed operator in Eq.~\eqref{eq:Ktilde}, first commute
one covariant momentum through $\hat U_y$:
\begin{align}
 \Pop_i [\hatxperp, \yperp]\,=\, [\hatxperp, \yperp] \mathrm{p}_i +  i\partial_{i}[\hatxperp, \yperp] \,+\, A_i [\hatxperp, \yperp].
\end{align}
The derivative of the transverse Wilson line is
\begin{align}
\partial_i[\hatxperp, \yperp]\,=\, iA_i(\xperp)[\hatxperp,\yperp]\,+
i\,\int_0^1\,d\alpha\,\alpha\,(\hatxperp - \yperp)_j\,[\hatxperp, \hatzetaperp{\alpha}] F_{ij}(\hatzetaperp{\alpha})[\hatzetaperp{\alpha}, \yperp]
,
\end{align}
where
$\hatzetaperp{\alpha}=\yperp+\alpha(\hatxperp-\yperp)$.  It follows that
\begin{align}
\label{eq: commute_transverse}
 \Pop_i  [\hatxperp, \yperp]\,=\,[\hatxperp, \yperp] p_i -  \int_0^1\,d\alpha\,\alpha\,(\hatxperp - \yperp)_j\,[\hatxperp, \hatzetaperp{\alpha}] F_{ij}(\hatzetaperp{\alpha})[\hatzetaperp{\alpha}, \yperp].
\end{align}
Equivalently,
\begin{align}
\label{eq: commutator_transverse}
 \Big[\mathrm{p}_i,   [\hatxperp, \yperp] \Big] \,=\, - A_i [\hatxperp, \yperp]  -  \int_0^1\,d\alpha\,\alpha\,(\hatxperp - \yperp)_j\,[\hatxperp, \hatzetaperp{\alpha}] F_{ij}(\hatzetaperp{\alpha})[\hatzetaperp{\alpha}, \yperp]\,.
\end{align}

Applying the second covariant momentum gives
\begin{align}
\Pop_{i} \left( \Pop_i  [\hatxperp, \yperp] \right)\,&=\, [\hatxperp, \yperp] \mathrm{p}_\perp^2
- \int_0^1\,d\alpha\,\alpha\,(\hatxperp - \yperp)_j\,[\hatxperp, \hatzetaperp{\alpha}] F_{ij}(\hatzetaperp{\alpha})[\hatzetaperp{\alpha}, \yperp] p_i \\ &
-  \int_0^1\,d\alpha\,\alpha\,\Pop_i(\hatxperp - \yperp)_j\,[\hatxperp, \hatzetaperp{\alpha}] F_{ij}(\hatzetaperp{\alpha})[\hatzetaperp{\alpha}, \yperp]
.
\end{align}
In the last term, $\mathrm p_i$ must be commuted to the right.  The
antisymmetry of $F_{ij}$ and
$[\mathrm p_i,\hat x_j]=\delta_{ij}$ eliminate the coordinate contribution.
After omitting terms that are simultaneously sub-eikonal and higher twist,
only the commutator with $F_{ij}(\hatzetaperp{\alpha})$ remains:
 \begin{align}
     [\mathrm{p}_i, F_{ij}(\hatzetaperp{\alpha})] = i \alpha D_i F_{ij}(\hatzetaperp{\alpha}) + \alpha[A_i, F_{ij}(\hatzetaperp{\alpha}) ] \,.
 \end{align}
Therefore,
\begin{align}
\Pop_{i} \left( \Pop_i  [\hatxperp, \yperp] \right)\,&=\, [\hatxperp, \yperp] \mathrm{p}_\perp^2
- 2  \int_0^1\,d\alpha\,\alpha\,(\hatxperp - \yperp)_j\,[\hatxperp, \hatzetaperp{\alpha}] F_{ij}(\hatzetaperp{\alpha})[\hatzetaperp{\alpha}, \yperp] p_i \na &
- i   \int_0^1\,d\alpha\,\alpha^2\,(\hatxperp - \yperp)_j\,[\hatxperp, \hatzetaperp{\alpha}] D_i F_{ij}(\hatzetaperp{\alpha})[\hatzetaperp{\alpha}, \yperp],
\end{align}
and hence
\begin{align}
[\yperp, \hatxperp] \Ptop^2 [\hatxperp, \yperp]&=\mathrm{p}_\perp^2  -  \int_0^1 d\alpha \,\alpha\Big(2 [\yperp, \hatzetaperp{\alpha}] (\hatxperp - \yperp)_jF_{ij}(\hatzetaperp{\alpha})[\hatzetaperp{\alpha}, \yperp]p_i +i \alpha [\yperp, \hatzetaperp{\alpha}] (\hatxperp - \yperp)_jD_{i}F_{ij}(\hatzetaperp{\alpha})[\hatzetaperp{\alpha}, \yperp]\Big) \na & 
= \mathrm{p}_\perp^2 + {\cal V}.
\end{align}
The transformed kernel is consequently
\begin{align}
\label{eq:trans_pert_expansion}
&\tilde{\mathcal{K}}_s(\xperp,\yperp)\,=\,(\xperp| e^{is U_y^{\dagger} \Ptop^2 U_y}|\yperp)\,=\,(\xperp| e^{is (\mathrm{p}_\perp^2 + {\cal V} )}|\yperp).
\end{align}
Applying the Duhamel expansion from Appendix~\ref{sec:duhamel} gives
\begin{align}
\tilde{\mathcal{K}}_s(\xperp,\yperp)\,&=\,(\xperp| e^{is \mathrm{p}_\perp^2}\,+\,i\int_0^s d\sigma\,e^{i(s-\sigma)\mathrm{p}_\perp^2 } \,{\cal V}\,e^{i\sigma \mathrm{p}_\perp^2 } + ...|\yperp) \na &
= (\xperp| e^{is \mathrm{p}_\perp^2}(1\,+\,i\int_0^s d\sigma\,e^{-i\sigma \mathrm{p}_\perp^2 } \,{\cal V}\,e^{i\sigma \mathrm{p}_\perp^2 } + ...)|\yperp).
\end{align}
Restoring the straight link yields
\begin{align}
\label{eq:exp_pert_expansion}
(\xperp|e^{i s \Ptop^2}|\yperp)=\mathcal{K}_s(\xperp,\yperp)\,=\,[\xperp,\yperp]\,(\xperp| e^{is \mathrm{p}_\perp^2}(1\,+\,i\int_0^s d\sigma\,e^{-i\sigma \mathrm{p}_\perp^2 } \,{\cal V}\,e^{i\sigma \mathrm{p}_\perp^2 } + ...)|\yperp)\,.
\end{align}
At the accuracy required by the eikonal and TMD limits, the term proportional
to $D_iF_{ij}$ contributes, whereas the other displayed term vanishes because
\begin{align}
   &(\xperp| e^{i(s-\sigma)\mathrm{p}_\perp^2 } \,  
   [\yperp, \hatzetaperp{\alpha}] (\hatxperp - \yperp)_jF_{ij}(\hatzetaperp{\alpha})[\hatzetaperp{\alpha}, \yperp]\mathrm{p}_i
   \,e^{i\sigma \mathrm{p}_\perp^2 } |\yperp)
    \na &=\int d^2 \zperp 
   (\xperp| e^{i(s-\sigma)\mathrm{p}_\perp^2 } \,  
   [\yperp, \hatzetaperp{\alpha}] (\hatxperp - \yperp)_jF_{ij}(\hatzetaperp{\alpha})[\hatzetaperp{\alpha}, \yperp] 
   |\zperp)(\zperp| \mathrm{p}_i
   \,e^{i\sigma \mathrm{p}_\perp^2 } |\yperp)
    \na &= \int d^2 \zperp 
   (\xperp| e^{i(s-\sigma)\mathrm{p}_\perp^2 }  |\zperp) \,  
   [\yperp,  \zetaperp{\alpha}] (\zperp - \yperp)_j F_{ij}(\zetaperp{\alpha})[\zetaperp{\alpha}, \yperp] 
   |\zperp)
   \underbrace{(\zperp| \mathrm{p}_i
   \,e^{i\sigma \mathrm{p}_\perp^2 } |\yperp)}_{\propto (\zperp - \yperp)_i} = 0 
\end{align}
vanishes by the antisymmetry of the field strength.  Therefore, at the
accuracy required here,
\begin{align}
\mathcal{K}_s(\xperp,\yperp)&=[\xperp,\yperp](\xperp| e^{is \mathrm{p}_\perp^2}\Big(1+\int_0^s d\sigma  \int_0^1\,d\alpha\, \alpha^2   e^{-i\sigma \mathrm{p}_\perp^2 }  [\yperp, \hatzetaperp{\alpha}] (\hatxperp - \yperp)_j D_{i}F_{ij}(\hatzetaperp{\alpha})[\hatzetaperp{\alpha}, \yperp]
\,e^{i\sigma \mathrm{p}_\perp^2 } + ...\Big)|\yperp)\,.
\label{eq:finalTE}
\end{align}
This is the required expansion of the transverse exponential.  The main text
also requires a covariant momentum inserted on its left:
\begin{align}
\mathcal{K}^{\rm 'L}_{s\,i}(\xperp,\yperp)\,=\,(\xperp|\Pop_i\,e^{i s \Ptop^2}|\yperp)\,.
\end{align}
The corresponding right-acting expression is obtained analogously.  For the
left-acting expression, we
generalize Eq.~\eqref{eq:Ktilde} by defining
\begin{align}
\tilde{\mathcal{K}}^{\rm 'L}_{s\,i}(\xperp,\yperp)\,=\,[\yperp,\xperp] \mathcal{K}^{\rm 'L}_{s\,i}(\xperp,\yperp) [\yperp,\yperp] = (\xperp| \hat{U}_y^{\dagger} \Pop_i e^{i s \Ptop^2} \hat{U}_y|\yperp)\,.
\end{align}
Using Eq.~\eqref{eq: commute_transverse},
\begin{align}
\tilde{\mathcal{K}}^{\rm 'L}_{s\,i}(\xperp,\yperp)\,&=\,(\xperp|  \mathrm{p}_i\hat{U}_y^{\dagger} e^{i s \Ptop^2} \hat{U}_y+\int_0^1\,d\alpha\,\alpha (\hatxperp - \yperp)_j[\yperp,\hatzetaperp{\alpha}]F_{ij}(\hatzetaperp{\alpha})[\hatzetaperp{\alpha},\yperp] \hat{U}_y^{\dagger} e^{i s \Ptop^2} \hat{U}_y |\yperp)\,
\na &
=\,(\xperp|  \mathrm{p}_i e^{i s \hat{U}_y^{\dagger}\Ptop^2\hat{U}_y} +\int_0^1\,d\alpha\,\alpha (\hatxperp - \yperp)_j[\yperp,\hatzetaperp{\alpha}]F_{ij}(\hatzetaperp{\alpha})[\hatzetaperp{\alpha},\yperp]  e^{i s \hat{U}_y^{\dagger}\Ptop^2\hat{U}_y}  |\yperp) \na &
=\,(\xperp|  \mathrm{p}_i e^{i s (\mathrm{p}_\perp^2 + \mathcal{V})} +\int_0^1\,d\alpha\,\alpha (\hatxperp - \yperp)_j[\yperp,\hatzetaperp{\alpha}]F_{ij}(\hatzetaperp{\alpha})[\hatzetaperp{\alpha},\yperp]  e^{i s (\mathrm{p}_\perp^2 + \mathcal{V})}  |\yperp)\,,
\end{align}
where the last equality uses Eq.~\eqref{eq:trans_pert_expansion}.  Expanding
with Eq.~\eqref{eq:exp_pert_expansion} gives
\begin{align}
\tilde{\mathcal{K}}^{\rm 'L}_{s\,i}(\xperp,\yperp)\,&=\,  (\xperp|  \mathrm{p}_i e^{is \mathrm{p}_\perp^2}(1\,+\,i\int_0^s d\sigma\,e^{-i\sigma \mathrm{p}_\perp^2 } \,{\cal V}\,e^{i\sigma \mathrm{p}_\perp^2 } + ...) +\int_0^1\,d\alpha\,\alpha (\hatxperp - \yperp)_j[\yperp,\hatzetaperp{\alpha}]F_{ij}(\hatzetaperp{\alpha})[\hatzetaperp{\alpha},\yperp]\,e^{is \mathrm{p}_\perp^2}\,\na &\,\times(1\,+\,i\int_0^s d\sigma\,e^{-i\sigma \mathrm{p}_\perp^2 } \,{\cal V}\,e^{i\sigma \mathrm{p}_\perp^2 } + ...)  |\yperp).
\end{align}
Retaining the terms required at the stated accuracy, we have
\begin{align}
\mathcal{K}^{\rm 'L}_{s\,i}(\xperp,\yperp)\,&=\, [\xperp,\yperp] (\xperp|  \mathrm{p}_i e^{is \mathrm{p}_\perp^2}(1\,+\,\int_0^s d\sigma\,  \int_0^1\,d\alpha\, \alpha^2 \, e^{-i\sigma \mathrm{p}_\perp^2 }  [\yperp, \hatzetaperp{\alpha}] (\hatxperp - \yperp)_j D_{i}F_{ij}(\hatzetaperp{\alpha})[\hatzetaperp{\alpha}, \yperp]
\,e^{i\sigma \mathrm{p}_\perp^2 }) \na &+\int_0^1\,d\alpha\,\alpha (\hatxperp - \yperp)_j[\yperp,\hatzetaperp{\alpha}]F_{ij}(\hatzetaperp{\alpha})[\hatzetaperp{\alpha},\yperp]\,e^{is \mathrm{p}_\perp^2}\, |\yperp)  
{}.
\end{align}
{\quad \bf \texorpdfstring{Comparison with Eq.~(70) of Ref.~\cite{Kar:2026vzk}}{Comparison with Eq. (70)}:}

Equation~\eqref{eq:finalTE} generalizes Eq.~(70) of
Ref.~\cite{Kar:2026vzk}.  To demonstrate the relation, insert intermediate
position eigenstates:
\begin{align*}
\mathcal{K}_s(\xperp,\yperp)\,&=\,[\xperp,\yperp]\,\Big((\xperp| e^{is \mathrm{p}_\perp^2}|\yperp)\,\na &+\,s\int_0^1 dt\,\int d^2 \zperp  \int_0^1\,d\alpha\, \alpha^2 \, (\xperp|e^{is(1-t) \mathrm{p}_\perp^2 }|\zperp)(\zperp|e^{ist\, \mathrm{p}_\perp^2 }|\yperp)  [\yperp, \zetaperp{\alpha}^z] (\zperp - \yperp)_j D_{i}F_{ij}( \zetaperp{\alpha}^z)[\zetaperp{\alpha}^z, \yperp]
\, + ...\Big)\,,
\end{align*}
where $t=\sigma/s$.  Using
\begin{align*}
(\xperp| e^{is \mathrm{p}_\perp^2}|\yperp)\,=\,\frac{i}{4\pi s}\,\exp\Big(-\frac{(\xperp-\yperp)^2}{4s}\Big),
\end{align*}
the product of free kernels can be rearranged as
\begin{align*}
(\xperp|e^{is(1-t) \mathrm{p}_\perp^2 }|\zperp)(\zperp|e^{is t \mathrm{p}_\perp^2 }|\yperp)\,&=\,\frac{i}{(4\pi)^2 s^2t (1-t)}\,\exp\Big(-\frac{(\xperp-\zperp)^2}{4 s(1-t)} -\frac{(\zperp-\yperp)^2}{4 st}\Big) \na &
= \frac{i}{(4\pi)^2 s \lambda}\,\exp\Big(-\frac{(\xperp-\yperp)^2}{4s} -\frac{w_\perp^2}{4\lambda}\Big) \na &
= (\xperp|e^{is p_\perp^2 }|\yperp)(\wperp|e^{i\lambda p_\perp^2 }|\zeroperp),
\end{align*}
where $\lambda=st(1-t)$ and
$\wperp=\zperp-t(\xperp-\yperp)-\yperp$.  Changing variables from $\zperp$ to
$\wperp$ gives
\begin{align*}
\mathcal{K}_s(\xperp,\yperp)\,&=\,[\xperp,\yperp]\,\Big((\xperp| e^{is \mathrm{p}_\perp^2}|\yperp)\,\na &+\,s\int_0^1 dt\,\int d^2 \wperp  \int_0^1\,d\alpha\, \alpha^2 \, (\xperp|e^{is \mathrm{p}_\perp^2 }|\yperp)(\wperp|e^{i\lambda \mathrm{p}_\perp^2 }|\zeroperp)  [\yperp, \zetaperp{\alpha}^w] (\wperp +  t (\xperp - \yperp))_j D_{i}F_{ij}(\zetaperp{\alpha}^w)[ \zetaperp{\alpha}^w, \yperp]
\, + ...\Big)\,,
\end{align*}
where
$\zetaperp{\alpha}^z=\zetaperp{\alpha}^w=\yperp+
\alpha[\wperp+t(\xperp-\yperp)]$.  The leading term in the $\wperp$ expansion
is
\begin{align*}
\mathcal{K}_s(\xperp,\yperp)\,&=\,[\xperp,\yperp]\,\Big((\xperp| e^{is \mathrm{p}_\perp^2}|\yperp)\,\na &+\,s\int_0^1 dt\,\int d^2 \wperp  \int_0^1\,d\alpha\, \alpha^2 \, (\xperp|e^{is \mathrm{p}_\perp^2 }|\yperp) (\wperp|e^{i\lambda \mathrm{p}_\perp^2 }|\zeroperp) [\yperp, \zetaperp{\alpha}^0] (\wperp +  t (\xperp - \yperp))_j D_{i}F_{ij}(\zetaperp{\alpha}^0)[\zetaperp{\alpha}^0, \yperp]
\, + ...\Big)\,.
\end{align*}
Performing the $\wperp$ integral gives
\begin{align*}
\mathcal{K}_s(\xperp,\yperp)\,&=\,[\xperp,\yperp]\,\Big((\xperp| e^{is \mathrm{p}_\perp^2}|\yperp)\,\na &+\,s\int_0^1 dt\,t\int_0^1\,d\alpha\, \alpha^2 \, (\xperp|e^{is \mathrm{p}_\perp^2 }|\yperp) [\yperp, \zetaperp{\alpha}^0] (\xperp - \yperp)_j D_{i}F_{ij}(\zetaperp{\alpha}^0)[\zetaperp{\alpha}^0, \yperp]
\, + ...\Big)\,,
\end{align*}
where $\zetaperp{\alpha}^0=\yperp+\alpha t(\xperp-\yperp)$; the term odd in
$\wperp$ vanishes.  Changing variables to $u=\alpha t$ and defining
$\zetaperp{u}=\yperp+u(\xperp-\yperp)$, we arrive at
\begin{align*}
\mathcal{K}_s(\xperp,\yperp)\,&=\,[\xperp,\yperp](\xperp| e^{is \mathrm{p}_\perp^2}|\yperp)\,\Big(1+\,s\int_0^1 du\,u\,\bar{u}\,(\xperp - \yperp)_j\, [\yperp, \zetaperp{u}]  D_{i}F_{ij}(\zetaperp{u})[\zetaperp{u}, \yperp]
\, + ...\Big)\,,
\end{align*}
which reproduces Eq.~(70) of Ref.~\cite{Kar:2026vzk}.

\section{\texorpdfstring{Fourier representation of the difference \(f(\xperp)-f(\yperp)\)}{Fourier representation of the difference f(x)-f(y)}}
\label{sec:fund_th_of_calc}

Throughout this appendix, let
\begin{equation}
    r^i \equiv x^i-y^i.
\end{equation}
Begin with the identity
\begin{equation}
    f(\xperp)
    = f(\yperp) + \left(f(\xperp) - f(\yperp)\right) = 
    f(\yperp)
    +
    \int d^2 \bperp\,
    \left[
        \delta^{(2)}(\xperp-\bperp)
        -
        \delta^{(2)}(\yperp-\bperp)
    \right]
    f(\bperp).
\end{equation}
Using the Fourier representation
\begin{equation}
    \delta^{(2)}(\zperp)
    =
    \int \frac{d^2 \kperp}{(2\pi)^2}\,
    e^{i \kperp\cdot \zperp},
\end{equation}
gives
\begin{equation}
    f(\xperp)
    =
    f(\yperp)
    +
    \int d^2 \bperp
    \int \frac{d^2 \kperp}{(2\pi)^2}
    \left[
        e^{i \kperp\cdot(\xperp-\bperp)}
        -
        e^{i \kperp\cdot(\yperp-\bperp)}
    \right]
    f(\bperp).
\end{equation}
Insert
\begin{align}
    1 = \frac{\rperp\cdot\kperp} {\rperp\cdot\kperp}  
\end{align}
in the integrand and replace the numerator $k_i$ by a derivative of the
exponentials.  Then
\begin{equation}
    f(\xperp)
    =
    f(\yperp)
    +
    i
    \int d^2 \bperp
    \int \frac{d^2 \kperp}{(2\pi)^2}\,
    r^i
    \frac{\partial}{\partial b^i}
    \left[
        \frac{
            e^{i \kperp\cdot(\xperp-\bperp)}
            -
            e^{i \kperp\cdot(\yperp-\bperp)}
        }{
            \rperp\cdot\kperp
        }
    \right]
    f(\bperp).
\end{equation}
Finally, integration by parts with respect to $\bperp$, with a vanishing
boundary term, gives
\begin{equation}
    f(\xperp)
    =
    f(\yperp)
    -
    i
    \int d^2 \bperp
    \int \frac{d^2 \kperp}{(2\pi)^2}\,
    \frac{
        e^{i \kperp\cdot(\xperp-\bperp)}
        -
        e^{i \kperp\cdot(\yperp-\bperp)}
    }{
        \rperp\cdot\kperp
    }
    r^i
    \,\partial_i f(\bperp)
    ,
  \label{eq:final-result2dtransporter}
\end{equation}
where
\begin{equation}
    r^i=x^i-y^i.
\end{equation}

The analogous one-dimensional identity is
\begin{equation}
  \begin{aligned}
  f(z^-)=f(y^-)
  &-i\int \dd b^-
    \int \frac{\dd  k^+}{2\pi}
    \frac{
      e^{i k^+(z^- - b^-)}
      -e^{i k^+ (y^- - b^-)}
    }
    {k^+ }
    \partial_- f(b^-).
  \end{aligned}
  \label{eq:final-result1dtransporter}
\end{equation}

\section{Reduction of the operator basis}
\label{app:reduction_F}
In this appendix we show that the background field operator appearing in the semi-infinite dipole contribution $i\mathcal{M}_0$, see Eq.~(\ref{eq:AmplitudeM2}), can be rewritten in terms of the background field operators appearing in the explicit field strength contributions (Fig.~\ref{fig:q_con} and Fig.~\ref{fig:aq_con}) in Eq.~(\ref{eq: quark_l_amp}) and Eq.~(\ref{eq: antiquark_l_amp}). Using Eq.~(\ref{eq:final-result2dtransporter}), we get a  field-strength insertion in Eq.~(\ref{eq:AmplitudeM2})
\begin{align}
i \mathcal{M}_{0,L}
 & =  -e \,\epsilon_-(q) \bar{u}(k_1)\,\gamma^-\,v(k_2)\, \frac{{ \delta(k_g^-) }}{2} \,  \int d^2 \mathbf{z}_{1\perp}  d^2 \mathbf{z}_{2\perp} \,e^{-i \mathbf{k}_{1\perp}\cdot\mathbf{z}_{1\perp} -i \mathbf{k}_{2\perp}\cdot\mathbf{z}_{2\perp}}\,
  \int { \bar d}^2 \mathbf{p}_{\perp} \;
     e^{i (\mathbf{z}_{1\perp} - \mathbf{z}_{2\perp}) \cdot\mathbf{p}_{\perp}  }
  \int_{-\infty}^{\infty} dy^-  \,\, e^{i k_{\mathbf{p}_\perp}^+   y^-}
  \na & \times
 \int d^2 \mathbf{b_\perp}\,\int \dhd^2 \mathbf{k_\perp}\,
 \frac{e^{i \mathbf{k_\perp}\cdot(\mathbf{z}_{1\perp} - \mathbf{b_\perp})} - e^{i \mathbf{k_\perp}\cdot(\mathbf{z}_{2\perp} - \mathbf{b_\perp})}}
 {(\mathbf{z}_{2\perp} - \mathbf{z}_{1\perp})\cdot\mathbf{k_\perp}}\,
 (\mathbf{z}_{2\perp} - \mathbf{z}_{1\perp})_i
 \Big(
 \partial_i([\infty,y^-]_{\mathbf{b_{\perp}}} [\mathbf{b_\perp},\mathbf{z}_{2\perp}]_{y^-} 
 \,[y^-,\infty]_{\mathbf{z}_{2\perp}})
 \na &
 - \partial_i([\infty,y^-]_{\mathbf{z}_{1\perp}} [\mathbf{z}_{1\perp},\mathbf{b_\perp}]_{y^-} 
 \,[y^-,\infty]_{\mathbf{b_\perp}})
 \Big)\,,
\end{align}
where we  dropped contributions independent of the background field. Computing the derivative of the staple, we obtain
\begin{align}
i \mathcal{M}_{0,L}
 & =  ie \,\epsilon_-(q) \bar{u}(k_1)\,\gamma^-\,v(k_2)\, \frac{{\delta(k_g^-) }}{2} \,  \int d^2 \mathbf{z}_{1\perp}  d^2 \mathbf{z}_{2\perp} \,e^{-i \mathbf{k}_{1\perp}\cdot\mathbf{z}_{1\perp} -i \mathbf{k}_{2\perp}\cdot\mathbf{z}_{2\perp}}\,
  \int { \bar d}^2 \mathbf{p}_{\perp} \;
     e^{i (\mathbf{z}_{1\perp} - \mathbf{z}_{2\perp}) \cdot\mathbf{p}_{\perp}  }
  \int_{-\infty}^{\infty} dy^-  \,\, e^{i k_{\mathbf{p}_\perp}^+   y^-}
  \na &\times
\int_{y^-}^{\infty} dz^-\int d^2 \mathbf{b_\perp}\int \dhd^2 \mathbf{k_\perp}
\frac{e^{i \mathbf{k_\perp}\cdot(\mathbf{z}_{1\perp} \!-\! \mathbf{b_\perp})} - e^{i \mathbf{k_\perp}\cdot(\mathbf{z}_{2\perp}\! -\! \mathbf{b_\perp})}}
{(\mathbf{z}_{2\perp} \!-\! \mathbf{z}_{1\perp})\cdot\mathbf{k_\perp}}
(\mathbf{z}_{2\perp}\! -\! \mathbf{z}_{1\perp})_{i}
\Big(
\bar F_{-i}(z^-,\mathbf{b_\perp})[\infty,y^-]_{\mathbf{b_{\perp}}} [\mathbf{b_\perp},\mathbf{z}_{2\perp}]_{y^-} [y^-,\infty]_{\mathbf{z}_{2\perp}}
 \na & 
 + [\infty,y^-]_{\mathbf{z}_{1\perp}} [\mathbf{z}_{1\perp},\mathbf{b_\perp}]_{y^-} 
[y^-,\infty]_{\mathbf{b_\perp}}\,\bar F_{-i}(z^-,\mathbf{b_\perp})
\Big)\,.
\end{align}
Finally evaluating the integral over $\mathbf{p}_\perp$,
\begin{align}
 \frac{2\pi}{q^-}\int { \bar d}^2 \mathbf{p}_{\perp} \,
     e^{i (\mathbf{z}_{1\perp} - \mathbf{z}_{2\perp}) \cdot\mathbf{p}_{\perp}  }
    \, e^{i k_{\mathbf{p}_\perp}^+   y^-}
\,=\,
\frac{i z \bar z}{y^-}\,
e^{-i q^+ y^-}\,
e^{-i \frac{z \bar z\,q^-}{2y^-}\,(\mathbf{z}_{1\perp} - \mathbf{z}_{2\perp})^2}\,
\end{align}
we get  
\begin{align}
i \mathcal{M}_{0,L}
 & =  -e \,\epsilon_-(q)\,z\bar z\,q^-\,\bar{u}(k_1)\,\gamma^-\,v(k_2)\, \frac{{\delta(k_g^-) }}{2} \,  \int d^2 \mathbf{z}_{1\perp}  d^2 \mathbf{z}_{2\perp} \,e^{-i \mathbf{k}_{1\perp}\cdot\mathbf{z}_{1\perp} -i \mathbf{k}_{2\perp}\cdot\mathbf{z}_{2\perp}}\,
  \int_{-\infty}^{\infty} dy^-  \,e^{-i q^+ y^-}\,
  e^{-i \frac{z \bar z\,q^-}{2y^-}\,(\mathbf{z}_{1\perp} - \mathbf{z}_{2\perp})^2}
  \na &
\times
\int_{y^-}^{\infty} dz^-\int d^2 \mathbf{b_\perp}\int \dhd^2 \mathbf{k_\perp}
\frac{e^{i \mathbf{k_\perp}\cdot(\mathbf{z}_{1\perp} \!-\! \mathbf{b_\perp})} - e^{i \mathbf{k_\perp}\cdot(\mathbf{z}_{2\perp}\! -\! \mathbf{b_\perp})}}
{(\mathbf{z}_{2\perp} \!-\! \mathbf{z}_{1\perp})\cdot\mathbf{k_\perp}}
(\mathbf{z}_{2\perp}\! -\! \mathbf{z}_{1\perp})_{i}
\Big(
\bar F_{-i}(z^-,\mathbf{b_\perp})[\infty,y^-]_{\mathbf{b_{\perp}}} [\mathbf{b_\perp},\mathbf{z}_{2\perp}]_{y^-} [y^-,\infty]_{\mathbf{z}_{2\perp}}
 \na &
 + [\infty,y^-]_{\mathbf{z}_{1\perp}} [\mathbf{z}_{1\perp},\mathbf{b_\perp}]_{y^-} 
 \,[y^-,\infty]_{\mathbf{b_\perp}}\,\bar F_{-i}(z^-,\mathbf{b_\perp})
\Big)\,.
\end{align}
This is an explicit representation in terms of the operators appearing in Eq.~(\ref{eq: quark_l_amp}) and Eq.~(\ref{eq: antiquark_l_amp}). 
\bibliography{refs.bib}

@article{delCastillo:2020omr,
    author = "del Castillo, Rafael F. and Echevarria, Miguel G. and Makris, Yiannis and Scimemi, Ignazio",
    title = "{TMD factorization for dijet and heavy-meson pair in DIS}",
    eprint = "2008.07531",
    archivePrefix = "arXiv",
    primaryClass = "hep-ph",
    doi = "10.1007/JHEP01(2021)088",
    journal = "JHEP",
    volume = "01",
    number = "2021",
    pages = "088",
    year = "2021"
}

@article{McLerran:1993ni,
    author = "McLerran, Larry D. and Venugopalan, Raju",
    title = "{Computing quark and gluon distribution functions for very large nuclei}",
    eprint = "hep-ph/9309289",
    archivePrefix = "arXiv",
    reportNumber = "TPI-MINN-93-44-T, NUC-MINN-93-24-T, HEP-UMN-TH-1220-93",
    doi = "10.1103/PhysRevD.49.2233",
    journal = "Phys. Rev. D",
    volume = "49",
    pages = "2233--2241",
    year = "1994"
}

@article{McLerran:1993ka,
    author = "McLerran, Larry D. and Venugopalan, Raju",
    title = "{Gluon distribution functions for very large nuclei at small transverse momentum}",
    eprint = "hep-ph/9311205",
    archivePrefix = "arXiv",
    reportNumber = "TPI-MINN-93-52-T, NUC-MINN-93-28-T, UMN-TH-1224-93",
    doi = "10.1103/PhysRevD.49.3352",
    journal = "Phys. Rev. D",
    volume = "49",
    pages = "3352--3355",
    year = "1994"
}

@article{Gelis:2010nm,
    author = "Gelis, Francois and Iancu, Edmond and Jalilian-Marian, Jamal and Venugopalan, Raju",
    title = "{The Color Glass Condensate}",
    eprint = "1002.0333",
    archivePrefix = "arXiv",
    primaryClass = "hep-ph",
    doi = "10.1146/annurev.nucl.010909.083629",
    journal = "Ann. Rev. Nucl. Part. Sci.",
    volume = "60",
    pages = "463--489",
    year = "2010"
}

@article{Dominguez:2011wm,
    author = "Dominguez, Fabio and Marquet, Cyrille and Xiao, Bo-Wen and Yuan, Feng",
    title = "{Universality of Unintegrated Gluon Distributions at small x}",
    eprint = "1101.0715",
    archivePrefix = "arXiv",
    primaryClass = "hep-ph",
    doi = "10.1103/PhysRevD.83.105005",
    journal = "Phys. Rev. D",
    volume = "83",
    pages = "105005",
    year = "2011"
}

@article{Dominguez:2011br,
    author = "Dominguez, Fabio and Qiu, Jian-Wei and Xiao, Bo-Wen and Yuan, Feng",
    title = "{On the linearly polarized gluon distributions in the color dipole model}",
    eprint = "1109.6293",
    archivePrefix = "arXiv",
    primaryClass = "hep-ph",
    doi = "10.1103/PhysRevD.85.045003",
    journal = "Phys. Rev. D",
    volume = "85",
    pages = "045003",
    year = "2012"
}

@article{Metz:2011wb,
    author = "Metz, Andreas and Zhou, Jian",
    title = "{Distribution of linearly polarized gluons inside a large nucleus}",
    eprint = "1105.1991",
    archivePrefix = "arXiv",
    primaryClass = "hep-ph",
    doi = "10.1103/PhysRevD.84.051503",
    journal = "Phys. Rev. D",
    volume = "84",
    pages = "051503",
    year = "2011"
}

@article{Dumitru:2015gaa,
    author = "Dumitru, Adrian and Lappi, Tuomas and Skokov, Vladimir",
    title = "{Distribution of Linearly Polarized Gluons and Elliptic Azimuthal Anisotropy in Deep Inelastic Scattering Dijet Production at High Energy}",
    eprint = "1508.04438",
    archivePrefix = "arXiv",
    primaryClass = "hep-ph",
    doi = "10.1103/PhysRevLett.115.252301",
    journal = "Phys. Rev. Lett.",
    volume = "115",
    number = "25",
    pages = "252301",
    year = "2015"
}

@article{Caucal:2023nci,
    author = {Caucal, Paul and Salazar, Farid and Schenke, Bj{\"o}rn and Stebel, Tomasz and Venugopalan, Raju},
    title = {{Back-to-back inclusive dijets in DIS at small x: gluon Weizs{\"a}cker-Williams distribution at NLO}},
    eprint = "2304.03304",
    archivePrefix = "arXiv",
    primaryClass = "hep-ph",
    doi = "10.1007/JHEP08(2023)062",
    journal = "JHEP",
    volume = "08",
    number = "2023",
    pages = "062",
    year = "2023"
}

@article{Caucal:2023fsf,
    author = {Caucal, Paul and Salazar, Farid and Schenke, Bj{\"o}rn and Stebel, Tomasz and Venugopalan, Raju},
    title = "{Back-to-Back Inclusive Dijets in Deep Inelastic Scattering at Small x: Complete NLO Results and Predictions}",
    eprint = "2308.00022",
    archivePrefix = "arXiv",
    primaryClass = "hep-ph",
    doi = "10.1103/PhysRevLett.132.081902",
    journal = "Phys. Rev. Lett.",
    volume = "132",
    number = "8",
    pages = "081902",
    year = "2024"
}

@article{Dumitru:2018kuw,
    author = "Dumitru, Adrian and Skokov, Vladimir and Ullrich, Thomas",
    title = {{Measuring the Weizs{\"a}cker-Williams distribution of linearly polarized gluons at an electron-ion collider through dijet azimuthal asymmetries}},
    eprint = "1809.02615",
    archivePrefix = "arXiv",
    primaryClass = "hep-ph",
    doi = "10.1103/PhysRevC.99.015204",
    journal = "Phys. Rev. C",
    volume = "99",
    number = "1",
    pages = "015204",
    year = "2019"
}

@article{Mantysaari:2019hkq,
    author = {M{\"a}ntysaari, Heikki and Mueller, Niklas and Salazar, Farid and Schenke, Bj{\"o}rn},
    title = "{Multigluon Correlations and Evidence of Saturation from Dijet Measurements at an Electron-Ion Collider}",
    eprint = "1912.05586",
    archivePrefix = "arXiv",
    primaryClass = "nucl-th",
    doi = "10.1103/PhysRevLett.124.112301",
    journal = "Phys. Rev. Lett.",
    volume = "124",
    number = "11",
    pages = "112301",
    year = "2020"
}

@article{AbdulKhalek:2021gbh,
    author = "Abdul Khalek, R. and others",
    title = "{Science Requirements and Detector Concepts for the Electron-Ion Collider}: {EIC Yellow Report}",
    eprint = "2103.05419",
    archivePrefix = "arXiv",
    primaryClass = "physics.ins-det",
    reportNumber = "BNL-220990-2021-FORE, JLAB-PHY-21-3198, LA-UR-21-20953",
    doi = "10.1016/j.nuclphysa.2022.122447",
    journal = "Nucl. Phys. A",
    volume = "1026",
    pages = "122447",
    year = "2022"
}

@article{Kotko:2015ura,
    author = "Kotko, P. and Kutak, K. and Marquet, C. and Petreska, E. and Sapeta, S. and van Hameren, A.",
    title = "{Improved TMD factorization for forward dijet production in dilute-dense hadronic collisions}",
    eprint = "1503.03421",
    archivePrefix = "arXiv",
    primaryClass = "hep-ph",
    reportNumber = "CERN-PH-TH-2015-045, CPHT-RR005.0315, IFJPAN-IV-2015-2",
    doi = "10.1007/JHEP09(2015)106",
    journal = "JHEP",
    volume = "09",
    number = "2015",
    pages = "106",
    year = "2015"
}

@article{Altinoluk:2019wyu,
    author = "Altinoluk, Tolga and Boussarie, Renaud",
    title = "{Low $x$ physics as an infinite twist (G)TMD framework: unravelling the origins of saturation}",
    eprint = "1902.07930",
    archivePrefix = "arXiv",
    primaryClass = "hep-ph",
    doi = "10.1007/JHEP10(2019)208",
    journal = "JHEP",
    volume = "10",
    number = "2019",
    pages = "208",
    year = "2019"
}

@article{Bury:2020ndc,
    author = "Bury, Marcin and van Hameren, Andreas and Kotko, Piotr and Kutak, Krzysztof",
    title = "{Forward trijet production in p-p and p-Pb collisions at LHC}",
    eprint = "2006.13175",
    archivePrefix = "arXiv",
    primaryClass = "hep-ph",
    doi = "10.1007/JHEP09(2020)175",
    journal = "JHEP",
    volume = "09",
    number = "2020",
    pages = "175",
    year = "2020"
}

@article{Altinoluk:2021ygv,
    author = "Altinoluk, Tolga and Marquet, Cyrille and Taels, Pieter",
    title = "{Low-x improved TMD approach to the lepto- and hadroproduction of a heavy-quark pair}",
    eprint = "2103.14495",
    archivePrefix = "arXiv",
    primaryClass = "hep-ph",
    doi = "10.1007/JHEP06(2021)085",
    journal = "JHEP",
    volume = "06",
    number = "2021",
    pages = "085",
    year = "2021"
}

@article{Fujii:2020bkl,
    author = "Fujii, Hirotsugu and Marquet, Cyrille and Watanabe, Kazuhiro",
    title = "{Comparison of improved TMD and CGC frameworks in forward quark dijet production}",
    eprint = "2006.16279",
    archivePrefix = "arXiv",
    primaryClass = "hep-ph",
    reportNumber = "JLAB-THY-20-3218",
    doi = "10.1007/JHEP12(2020)181",
    journal = "JHEP",
    volume = "12",
    number = "2020",
    pages = "181",
    year = "2020"
}

@article{Kutak:2021kaw,
    author = "Kutak, Krzysztof",
    title = "{Small-x Improved Transverse Momentum Dependent factorization and some of its recent applications}",
    eprint = "2111.02286",
    archivePrefix = "arXiv",
    primaryClass = "hep-ph",
    doi = "10.22323/1.398.0412",
    journal = "PoS",
    volume = "EPS-HEP2021",
    pages = "412",
    year = "2022"
}

@article{Ganguli:2023joy,
    author = "Ganguli, Ishita and van Hameren, Andreas and Kotko, Piotr and Kutak, Krzysztof",
    title = "{Forward $\gamma $+jet production in proton-proton and proton-lead collisions at LHC within the FoCal calorimeter acceptance}",
    eprint = "2306.04706",
    archivePrefix = "arXiv",
    primaryClass = "hep-ph",
    doi = "10.1140/epjc/s10052-023-12043-3",
    journal = "Eur. Phys. J. C",
    volume = "83",
    number = "9",
    pages = "868",
    year = "2023"
}

@article{Boussarie:2021ybe,
    author = {Boussarie, Renaud and M{\"a}ntysaari, Heikki and Salazar, Farid and Schenke, Bj{\"o}rn},
    title = "{The importance of kinematic twists and genuine saturation effects in dijet production at the Electron-Ion Collider}",
    eprint = "2106.11301",
    archivePrefix = "arXiv",
    primaryClass = "hep-ph",
    doi = "10.1007/JHEP09(2021)178",
    journal = "JHEP",
    volume = "09",
    number = "2021",
    pages = "178",
    year = "2021"
}

@article{Altinoluk:2022jkk,
    author = "Altinoluk, Tolga and Beuf, Guillaume and Czajka, Alina and Tymowska, Arantxa",
    title = "{DIS dijet production at next-to-eikonal accuracy in the CGC}",
    eprint = "2212.10484",
    archivePrefix = "arXiv",
    primaryClass = "hep-ph",
    doi = "10.1103/PhysRevD.107.074016",
    journal = "Phys. Rev. D",
    volume = "107",
    number = "7",
    pages = "074016",
    year = "2023"
}

@article{Altinoluk:2024zom,
    author = "Altinoluk, Tolga and Beuf, Guillaume and Czajka, Alina and Marquet, Cyrille",
    title = "{Back-to-back dijet production in DIS at next-to-eikonal accuracy and twist-3 gluon TMDs}",
    eprint = "2410.00612",
    archivePrefix = "arXiv",
    primaryClass = "hep-ph",
    doi = "10.1103/PhysRevD.111.014010",
    journal = "Phys. Rev. D",
    volume = "111",
    number = "1",
    pages = "014010",
    year = "2025"
}

@article{Agostini:2024xqs,
    author = "Agostini, Pedro and Altinoluk, Tolga and Armesto, N{\'e}stor",
    title = "{Next-to-eikonal corrections to dijet production in Deep Inelastic Scattering in the dilute limit of the Color Glass Condensate}",
    eprint = "2403.04603",
    archivePrefix = "arXiv",
    primaryClass = "hep-ph",
    doi = "10.1007/JHEP07(2024)137",
    journal = "JHEP",
    volume = "07",
    number = "2024",
    pages = "137",
    year = "2024"
}

@misc{Armesto:2026qwh,
    author = "Armesto, N{\'e}stor and Dom{\'\i}nguez, Fabio and Romero, Adri{\'a}n",
    title = "{Non-eikonal corrections to dijet production in DIS}",
    eprint = "2603.26526",
    archivePrefix = "arXiv",
    primaryClass = "hep-ph",
    month = "3",
    year = "2026"
}

@inproceedings{Altinoluk:2026gdc,
    author = "Altinoluk, Tolga and Beuf, Guillaume and Czajka, Alina and Marquet, Cyrille",
    title = "{Back-to-back dijet production in DIS with finite-energy corrections and twist-3 gluon TMDs}",
    eprint = "2603.09960",
    archivePrefix = "arXiv",
    primaryClass = "hep-ph",
    month = "3",
    year = "2026"
}

@misc{Mukherjee:2026cte,
    author = "Mukherjee, Swagato and Skokov, Vladimir V. and Tarasov, Andrey and Tiwari, Shaswat and Yao, Fei",
    title = "{Back-to-back dijet production in DIS at arbitrary Bjorken-x: TMD gluon distributions to twist-3 accuracy}",
    eprint = "2602.15137",
    archivePrefix = "arXiv",
    primaryClass = "hep-ph",
    month = "2",
    year = "2026"
}

@misc{Mukherjee:2026six,
    author = "Mukherjee, Swagato and Skokov, Vladimir. V. and Tarasov, Andrey and Tiwari, Shaswat and Yao, Fei",
    title = "{Back-to-back dijet production in DIS at arbitrary Bjorken-x: TMD quark distributions to twist-3 accuracy}",
    eprint = "2607.12268",
    archivePrefix = "arXiv",
    primaryClass = "hep-ph",
    month = "7",
    year = "2026"
}

@misc{Kar:2026vzk,
    author = "Kar, Tiyasa and Tarasov, Andrey and Skokov, Vladimir V.",
    title = "{DIS dijet production in Background Field Approach: General formalism and methods}",
    eprint = "2603.08805",
    archivePrefix = "arXiv",
    primaryClass = "hep-ph",
    month = "3",
    year = "2026"
}

@article{CaucalEtAl2021DijetNLO,
  author        = {Caucal, Paul and Salazar, Farid and
                   Venugopalan, Raju},
  title         = {Dijet impact factor in {DIS} at next-to-leading
                   order in the Color Glass Condensate},
  journal       = {JHEP},
  volume        = {11},
  number = "2021",
  pages         = {222},
  year          = {2021},
  doi           = {10.1007/JHEP11(2021)222},
  eprint        = {2108.06347},
  archivePrefix = {arXiv},
  primaryClass  = {hep-ph}
}

@article{CaucalEtAl2022BackToBackNLO,
  author        = {Caucal, Paul and Salazar, Farid and
                   Schenke, Bj{\"o}rn and Venugopalan, Raju},
  title         = {Back-to-back inclusive dijets in {DIS} at small
                   $x$: Sudakov suppression and gluon saturation
                   at {NLO}},
  journal       = {JHEP},
  volume        = {11},
  number = "2022",
  pages         = {169},
  year          = {2022},
  doi           = {10.1007/JHEP11(2022)169},
  eprint        = {2208.13872},
  archivePrefix = {arXiv},
  primaryClass  = {hep-ph}
}

@article{BoussarieMehtarTani2022NovelUGD,
  author        = {Boussarie, Renaud and Mehtar-Tani, Yacine},
  title         = {A novel formulation of the unintegrated gluon
                   distribution for {DIS}},
  journal       = {Phys. Lett. B},
  volume        = {831},
  pages         = {137125},
  year          = {2022},
  doi           = {10.1016/j.physletb.2022.137125},
  eprint        = {2006.14569},
  archivePrefix = {arXiv},
  primaryClass  = {hep-ph}
}

@article{BoussarieMehtarTani2022PartialTwist,
  author        = {Boussarie, Renaud and Mehtar-Tani, Yacine},
  title         = {Gluon-mediated inclusive Deep Inelastic Scattering
                   from Regge to Bjorken kinematics},
  journal       = {JHEP},
  volume        = {07},
  number = "2022",
  pages         = {080},
  year          = {2022},
  doi           = {10.1007/JHEP07(2022)080},
  eprint        = {2112.01412},
  archivePrefix = {arXiv},
  primaryClass  = {hep-ph}
}

@article{BoussarieMehtarTani2024ComptonPTE,
  author        = {Boussarie, Renaud and Mehtar-Tani, Yacine},
  title         = {Low and moderate $x$ gluon contribution to exclusive
                   Compton scattering processes},
  journal       = {JHEP},
  volume        = {10},
  number = "2024",
  pages         = {056},
  year          = {2024},
  doi           = {10.1007/JHEP10(2024)056},
  eprint        = {2309.16576},
  archivePrefix = {arXiv},
  primaryClass  = {hep-ph}
}

\end{document}